# Universal Function Approximation via Diffractive Optical Processors: Physical Limits, Error Bounds, and Learnability

Md Sadman Sakib Rahman[1,2,3], Che-Yung Shen[1,2,3] and Aydogan Ozcan[1,2,3,*]

[1] Electrical and Computer Engineering Department, University of California, Los Angeles, CA, 90095, USA

[2] Bioengineering Department, University of California, Los Angeles, CA, 90095, USA

[3] California NanoSystems Institute (CNSI), University of California, Los Angeles, CA, 90095, USA

* ozcan@ucla.edu

## Abstract

Universal approximation theorems establish the expressive power of neural network architectures, but they do not address the physical constraints governing approximation accuracy, learnability, scalability, or energy efficiency in analog optical implementations. Recent diffractive optical processors have demonstrated massively parallel approximation of nonlinear functions through phase-encoded wavefronts and passive linear propagation, establishing a compelling paradigm for optical nonlinear computation. Here, we present a unified theoretical framework connecting classical universal approximation theory, Fourier-feature approximation, and diffractive optical processors. We show that phase-encoded diffractive processors implement finite Fourier-feature expansions whose mathematical completeness follows from Fourier/Stone-Weierstrass arguments, while their physical realizability is governed by finite coefficient synthesis through optimized spatially varying coherent point-spread functions (PSFs). Our analyses derive approximation-error bounds that separate Fourier truncation, PSF-synthesis, input phase error, optical hardware, readout, and noise contributions; establish scaling relationships linking approximation complexity to optical degrees of freedom and input/output space-bandwidth products; derive photon-budget and throughput limits imposed by photon statistics; formulate finite-class statistical learnability bounds for phase-quantized diffractive function approximators; and analyze the impact of spatially incoherent illumination. We further analyze coherent optical cascadability and show that quadratic feature expansion through coherent mixing and optical readout provides a mechanism for enhanced representation while remaining fundamentally distinct from the depth-separation results established for digital neural networks. Our analyses provide a rigorous theoretical foundation for diffractive nonlinear function approximation and establish quantitative relationships among mathematical expressivity, optical hardware resources, statistical learning, and physical performance limits, thereby offering general design principles for large-scale analog optical computing systems.

## Introduction

Universal function approximation (UFA) theorems provide a foundational mathematical language for understanding why neural networks can represent extraordinarily broad classes of functions. They establish that sufficiently expressive neural-network families are dense in broad classes of target functions, including continuous functions on compact domains and measurable functions under appropriate metrics[1–9]. These results, however, are fundamentally existence theorems: they establish representational capability but do not address the physical or statistical considerations[1,10] that determine whether a universal approximator can be efficiently realized, trained, generalized, or implemented under finite computational and/or hardware resources.

Recent advances in diffractive optical processors have demonstrated that passive linear optical systems can perform massively parallel approximation of nonlinear functions by encoding the input variables into the phase of an optical wavefront and transforming these encoded fields through optimized diffractive structures with spatially varying coherent point-spread functions (PSFs)[11]. Under this wavefront encoding, a diffractive processor implements a finite Fourier-feature expansion in optical hardware, enabling the simultaneous approximation of a large number of nonlinear functions at diffraction-limited spatial density. For bandlimited target functions, or functions represented through finite Fourier truncations, this architecture provides a physically scalable realization of massively parallel nonlinear function approximation.

Here, we report a unified theoretical framework that places diffractive optical function approximators within the broader context of classical universal approximation theory, Fourier-feature approximation, harmonic analysis, and statistical learning. In addition to establishing representability, we quantify the physical and statistical factors governing the practical realization of diffractive function approximators. Specifically, we (1) derive approximation-error bounds that separate error terms due to Fourier-truncation, coefficient-synthesis, input phase variations and hardware-induced errors; (2) establish scaling relationships linking approximation complexity to optical degrees of freedom and input/output space-bandwidth products; (3) analyze the energy, throughput, and precision limits imposed by diffraction efficiency and photon statistics; (4) derive finite-class statistical learnability bounds for phase-quantized diffractive function approximators; (5) examine coherent optical cascadability in relation to feature expansion and depth-enabled improved representation; and (6) analyze the capabilities/features of spatially incoherent diffractive function approximators. Together, these results provide a theoretical foundation that connects mathematical expressivity with optical hardware constraints, statistical learning, and physical performance limits, thereby establishing general design principles for future large-scale analog diffractive optical computing systems.

We emphasize that the presented framework represents a fundamental departure from the conventional Fourier optics paradigm. Although both employ optical propagation and Fourier representations, the Fourier quantities appearing in this work have fundamentally different meanings. In classical Fourier optics, the Fourier variables describe the spatial-frequency or angular-spectrum content of propagating electromagnetic waves, and optical propagation manipulates these physical spatial-frequency components through diffraction and spatial filtering. By contrast, the Fourier variables employed here are computational basis indices associated with the mathematical variables of the nonlinear functions being approximated. They do not represent optical wave vectors, propagation angles, diffraction orders, or the spatial-frequency spectrum of the propagating electromagnetic field. Instead, these computational Fourier frequencies are prescribed during the input phase-encoding stage to construct a finite Fourier representation of the target functions before optical propagation begins. The diffractive processor then performs a fixed linear optical transformation whose optimized spatially varying coherent point-spread functions synthesize the corresponding Fourier coefficients. Consequently, the Fourier representation belongs to the abstract function space of the encoded variables. Furthermore, the Fourier basis employed in this framework is defined over an arbitrary mathematical variable $\mathbf{a} \in \mathbb{R}^D$, where the dimensionality $D$ is determined by the function-approximation problem rather than by physical space and may therefore greatly exceed three ($D \gg 3$). Thus, while the optical system propagates ordinary three-dimensional electromagnetic fields, it simultaneously realizes programmable Fourier expansions in abstract high-dimensional function spaces. The proposed architecture should therefore be viewed not as a generalization of classical Fourier optics, but rather as a programmable physical realization of high-dimensional Fourier-function approximation in which coherent (or partially coherent or incoherent[12]) wave propagation serves as the hardware for implementing learned expansions in abstract computational spaces.

## Results

### Diffractive Optical Processors as Fourier-Feature Universal Function Approximators

Classical UFA theorems state that certain families of finite neural network sums are dense in target function spaces. In the standard single-hidden-layer form, for an input vector $\boldsymbol{x} \in \mathbb{R}^n$, we have[3,4,13]:

$$G(\boldsymbol{x}) = \sum_{j=1}^{N_\sigma} c_j \ \sigma\big(\boldsymbol{y}_{\boldsymbol{j}}^{\boldsymbol{T}} \boldsymbol{x} + \theta_j\big), \qquad c_j, \theta_j \in \mathbb{R},\ \boldsymbol{y}_{\boldsymbol{j}} \in \mathbb{R}^n \quad \textbf{Eq. (1)}$$

Here, $G(\boldsymbol{x})$ is the output function to be approximated, $c_j$ and $\theta_j$ are scalar output weights and biases, $\boldsymbol{y}_{\boldsymbol{j}}$ is a vector of input weights, $\sigma$ is a scalar activation function, and $N_\sigma$ refers to the number of hidden nonlinear units. Cybenko proved uniform density in $C([0,1]^n)$ for continuous discriminatory functions, with continuous sigmoidal activation functions as the central neural-network backbone[3]. Hornik, Stinchcombe, and White established broad approximation results for multilayer feedforward networks, including the approximation of Borel-measurable functions under appropriate metrics[4]. Later work expanded the activation-function condition: for standard feedforward networks with locally bounded piecewise-continuous activations, non-polynomiality of the activation is the key condition for density[5]. Unbounded activations, such as ReLU (Rectified Linear Unit), also ensure universal approximation; this follows from the broader non-polynomial activation literature, while Sonoda and Murata provide a constructive ridgelet-transform analysis for such activation functions[9]. Width- and depth-specific ReLU universality results, such as Lu et al., also provide additional representational statements[6].

Recently, we reported that a diffractive optical processor can perform multiplexed approximation of a large set of nonlinear functions[11] executed in parallel at its output plane using phase encoding and spatially varying PSFs optimized via supervised learning. In this architecture, the intensity-based demonstrations use square-law detection, whereas the complex-valued and all-optically cascadable demonstrations preserve the coherent output field and therefore do not require intensity readout at an intermediate stage. Stated differently, the nonlinear function-approximation capability of a diffractive optical processor does ***not*** necessarily rely on intensity detection and has been demonstrated to be cascadable by using the complex optical field to represent a desired complex-valued nonlinear function for each output aperture/channel. Such a diffractive optical processor and the function approximation architecture behind it rely on Fourier-based function approximation using finite sums of complex exponentials, typically on compact or periodic domains, or by truncating a Fourier representation. Under a desired and programmed input phase encoding, the physical diffractive processor implements a finite Fourier-feature expansion in optical hardware; see Fig. 1.

In a coherent diffractive optical network, the physical propagation of the wave through the diffractive volume is linear in the optical field. The nonlinearity with respect to the variables being computed enters through the input phase encoding. For a $D$-dimensional input variable $\mathbf{a} = (a_1, \dots, a_D)$, one can assign an input-pixel-dependent frequency vector $\boldsymbol{\alpha}_p \in \mathbb{R}^D$ and spatially encode:

$$u_{\text{in}}(p; \mathbf{a}) = \exp\big(j2\pi\, \boldsymbol{\alpha}_p \cdot \mathbf{a}\big) \quad \textbf{Eq. (2)}$$

The optimized diffractive processor supplies complex-valued spatially varying connection coefficients $\hat{F}(k; p)$ between the input pixels $p$ and the output channels $k$. The complex field that is processed by the diffractive network at the output channel $k$ can be written as:

$$\tilde{f}_k(\mathbf{a}) \equiv u_{\text{out}}(k; \mathbf{a}) = \sum_{p=1}^{N_p} \hat{F}\,(k; p) \exp\big(j2\pi\, \boldsymbol{\alpha}_p \cdot \mathbf{a}\big) \quad \textbf{Eq. (3)}$$

This is a finite $D$-dimensional Fourier-feature model. The basis functions are fixed by the input encoding; the trainable optical processor implements the complex-valued coefficient matrix and the physical wave routing needed to deliver those coefficients at the output plane, executing in parallel $N_f$ distinct nonlinear functions $\tilde{f}_k(\mathbf{a})$ at the output, $k = 1,2, \dots, N_f$. Because the diffractive optical processor is passive, $\hat{F}(k; p)$ in Eq. (3) should be interpreted as an effective, calibrated complex-valued coefficient matrix. Its normalization accounts for the chosen input/output field scaling, output diffraction efficiency, optical losses, and readout gain; the corresponding throughput and noise penalties are treated separately in our subsequent analyses in the next sections.

If each target (ground truth) function $f_k(\mathbf{a})$ can be accurately replaced by a finite Fourier approximation $f_{k,\Lambda}$ over a frequency set $\Lambda = \{\boldsymbol{\alpha}_p\}_{p=1}^{N_p}$ which in general includes negative/positive frequencies and DC, then we have:

$$f_k(\mathbf{a}) \approx f_{k,\Lambda}(\mathbf{a}) = \sum_{\boldsymbol{\alpha}_p \in \Lambda} F_k\left(\boldsymbol{\alpha}_p\right) \exp\left(j2\pi\, \boldsymbol{\alpha}_p \cdot \mathbf{a}\right) \quad \mathbf{Eq. (4)}$$

A diffractive optical processor that is optimized (using e.g., supervised learning) to realize $\hat{F}(k;p) \approx F_k(\boldsymbol{\alpha}_p)$ computes these finite Fourier approximations for many target functions in parallel[11], for the simultaneous approximation of $N_f$ nonlinear functions (i.e., $\tilde{f}_k(\mathbf{a}) \approx f_k(\mathbf{a})$ for $k = 1,2,\dots,N_f$) at its output plane; see Figs. 1-2. The accuracy of $\hat{F}(k;p) \approx F_k(\boldsymbol{\alpha}_p)$ can be connected to the arbitrary complex-valued linear-transformation synthesis capabilities of diffractive networks[14–19]. For a diffractive system implementing a complex-valued linear transformation from an input field vector of dimension $N_i$ (corresponding to the input space-bandwidth product of the diffractive processor) to an output field vector of dimension $N_o$ (corresponding to the output space-bandwidth product of the same diffractive processor), our former work showed that negligible transformation error is achieved when the number ($N$) of optimized complex-valued diffractive features satisfies $N \gtrsim N_i N_o$; this condition was shown to increase to $N \gtrsim 2N_i N_o$ for $N$ phase-only trainable diffractive features[14–19]. We also validated the same diffractive synthesis framework on unitary, nonunitary and noninvertible transformations[14]. In the presented diffractive function-approximation setting, the input vector is the $N_p$-dimensional Fourier-feature vector and the output vector contains $N_f$ output function channels. Thus, the corresponding coefficient matrix $\hat{F}(k;p) \in \mathbb{C}^{N_f \times N_p}$ has $N_f N_p$ complex-valued entries, consistent with the scaling $N \gtrsim 2N_i N_o = 2N_p N_f$ for a diffractive optical processor with $N$ phase-only diffractive features that are trainable.

To shed more light on this, Fig. 2 establishes the scalability of the diffractive architecture for optical field-based approximation of a large number of complex-valued nonlinear functions. As the number of output nonlinear functions $N_f$ increases, both the spread and the maximum of the error distribution increase, reflecting the growing complexity of the task. Nevertheless, the errors remain remarkably low across all designs, including for $N_f \approx 10^5$, underscoring the inherent scalability of the diffractive processor architecture for massively parallel execution of nonlinear functions. To further illustrate this performance, Fig. 2b presents specific examples of optical field-based computed functions from the two larger-scale designs: $N_f = 10^4$ and $N_f \approx 10^5$. In each case, the real and imaginary parts of the target nonlinear functions are shown alongside the corresponding diffractive approximation. Figure 2b also includes the functions with the largest approximation error from each design, revealing very good agreement between the target and the diffractive output even in these worst cases.

The complex exponential feature $\exp\left(j2\pi\boldsymbol{\alpha}_p \cdot \mathbf{a}\right)$ is not a trainable ridge activation of the form $\sigma\left(\boldsymbol{y}_j^T \boldsymbol{x} + \theta_j\right)$ as in the usual single-hidden-layer neural-network UFA theorem. Following Eq. (3), it is a Fourier component indexed by $\boldsymbol{\alpha}_p$. Mathematical completeness of the associated Fourier-feature family follows from Fourier/Stone-Weierstrass arguments in settings where such exponentials form a separating algebra, most cleanly on compact periodic domains. When the complex exponentials under consideration generate a unital trigonometric algebra that separates points on the compact domain and is closed under complex conjugation, finite linear combinations of these exponentials are dense in the relevant space of continuous functions by the Stone-Weierstrass theorem[20–22]. While a single finite frequency grid is not dense, the union of finite Fourier-feature spaces over increasingly rich frequency sets is dense under the usual Fourier/Stone-Weierstrass assumptions. A diffractive processor with a finite $N_p$ at the input plane realizes one finite member of this family through input phase encoding, and if its spatially varying PSFs synthesize the required coefficients accurately, i.e., $\hat{F}(k;p) \approx F_k(\boldsymbol{\alpha}_p)$, it approximates the corresponding finite Fourier truncation. Stated differently, for any fixed finite input frequency set and fixed diffractive geometry, the optical processor realizes only a finite-dimensional approximation class; universality in this context refers to an architecture family with increasingly rich Fourier-feature sets and physically realizable spatially varying PSF-synthesis capability of the diffractive network. In our diffractive function approximator architecture[11], the direct approximation argument is based on bandlimited functions or finite Fourier truncations. For nonperiodic functions on a finite interval, the Fourier representation should be understood after choosing a domain, scaling, and extension convention. Discontinuities or derivative mismatches at the boundary of the periodic extension can produce Gibbs-type or edge artifacts, as reported in our former work[11], which can be reduced by, e.g., increasing bandwidth, smoothing or windowing or enlarging the approximation interval.

Therefore, the complex output field $\tilde{f}_k(\mathbf{a})$ of a diffractive processor can represent complex-valued Fourier sums if the output amplitude and phase are measured, interferometrically recovered, or coherently fed to another cascaded optical processor, as shown in our prior work[11]. Direct intensity detection, instead, measures $\left|\tilde{f}_k(\mathbf{a})\right|^2$, which is nonnegative and discards the phase unless a phase-sensitive measurement scheme is used. Therefore, arbitrary signed or complex-valued function approximation requires an explicit readout convention, such as differential channels, baseline offsets, separate real/imaginary channels, or coherent field transfer of $\tilde{f}_k(\mathbf{a})$ onto a cascaded optical processor for subsequent processing, which will be further detailed/discussed in our analyses below.

### Diffractive Function Approximation Error Bounds and Physical Scaling Laws

In our notation, $\tilde{f}_k$ denotes the complex optical field at the output of an optimized diffractive optical processor (i.e., the function approximator), while $\hat{f}_k$ denotes the final measured or decoded field after readout. Let $K \subset \mathbb{R}^D$ be the compact domain of interest and let $\Lambda = \{\boldsymbol{\alpha}_p\}_{p=1}^{N_p}$ be the frequency set implemented by the input pixels. Following the earlier notation in the previous section, we define the Fourier-truncated target function as $f_{k,\Lambda}(\mathbf{a}) = \sum_{\boldsymbol{\alpha}_p \in \Lambda} F_k\left(\boldsymbol{\alpha}_p\right) \exp\left(j2\pi\, \boldsymbol{\alpha}_p \cdot \mathbf{a}\right)$ and the complex field computed by the physical diffractive processor as $\tilde{f}_k(\mathbf{a}) = \sum_{\boldsymbol{\alpha}_p \in \Lambda} \hat{F}\,(k;p) \exp\left(j2\pi\, \boldsymbol{\alpha}_p \cdot \mathbf{a}\right)$.

To evaluate the error bound due to the optimization of the spatially varying PSFs of a diffractive optical processor, let us define $c_{k,p} = F_k(\boldsymbol{\alpha}_p) - \hat{F}(k;p)$. Then, we have:

$$f_{k,\Lambda}(\mathbf{a}) - \tilde{f}_k(\mathbf{a}) = \sum_{\boldsymbol{\alpha}_p \in \Lambda} c_{k,p} \exp\left(j2\pi\, \boldsymbol{\alpha}_p \cdot \mathbf{a}\right).$$

$$\| f_{k,\Lambda}(\mathbf{a}) - \tilde{f}_k(\mathbf{a}) \|_{\infty,K} = \sup_{\mathbf{a}\in K} \left| \sum_{\boldsymbol{\alpha}_p \in \Lambda} c_{k,p} \exp\left(j2\pi\, \boldsymbol{\alpha}_p \cdot \mathbf{a}\right) \right|$$

where the maximum error in the diffractive function approximation of $f_{k,\Lambda}(\mathbf{a})$ is expressed through the supremum (sup) norm over the domain of interest, $K \subset \mathbb{R}^D$. Using the triangle inequality, we have:

$$\left| \sum_{\boldsymbol{\alpha}_p \in \Lambda} c_{k,p} \exp\left(j2\pi\, \boldsymbol{\alpha}_p \cdot \mathbf{a}\right) \right| \le \sum_{\boldsymbol{\alpha}_p \in \Lambda} |c_{k,p}|,$$

which yields,

$$\| f_{k,\Lambda}(\mathbf{a}) - \tilde{f}_k(\mathbf{a}) \|_{\infty,K} \le \sum_{\boldsymbol{\alpha}_p \in \Lambda} \left| F_k(\boldsymbol{\alpha}_p) - \hat{F}(k;p) \right| = E_{\mathrm{PSF},k} \quad \textbf{Eq. (5)}$$

Here $E_{\mathrm{PSF},k} = \sum_{\boldsymbol{\alpha}_p \in \Lambda} \left| F_k(\boldsymbol{\alpha}_p) - \hat{F}(k;p) \right|$ represents the PSF approximation error of the diffractive processor due to, e.g., local minima in its optimization and/or limited diffractive degrees of freedom in the optical design. As discussed earlier in relationship to Eq. (3), the definition of $\hat{F}(k;p)$ assumes a scalar factor applied to the learned PSFs to account for diffraction-induced losses, and this complex field-based function approximation error bound is correct after the target has been reduced to the same finite frequency grid used by the optical input.

To shed more light on Eq. (5), in Fig. 3 we show that $E_{f,k} = \| f_{k,\Lambda}(\mathbf{a}) - \tilde{f}_k(\mathbf{a}) \|_{\infty,K}$ consistently remains below the corresponding $E_{\mathrm{PSF},k}$; similarly, Fig. 4 shows that the $(E_{\mathrm{PSF},k}, E_{f,k})$ points for the individual nonlinear functions lie below the line $E_f = E_{\mathrm{PSF}}$. Together, these results are consistent with the error bound reported in Eq. (5). It is also noteworthy from Fig. 3 that the errors can be substantial when $N < 2N_pN_f$, but become negligible once $N$ exceeds $2N_pN_f$, supporting the diffractive-feature scaling requirement discussed above. Furthermore, distributing the same number of trainable diffractive features over a deeper architecture with more diffractive layers (i.e., larger $K_L$) reduces both the PSF-synthesis and function-approximation errors, as also confirmed in Figs. 3-4.

Equation (5) is a conservative uniform bound based on the row-wise $\ell_1$ PSF error. Other transformation-error metrics, such as row-wise $\ell_2$ PSF error, can be related to this bound through:

$$\sqrt{\sum_{\boldsymbol{\alpha}_p \in \Lambda} \left|F_k(\boldsymbol{\alpha}_p) - \hat{F}(k;p)\right|^2} \leq E_{\text{PSF},k} = \sum_{\boldsymbol{\alpha}_p \in \Lambda} \left|F_k(\boldsymbol{\alpha}_p) - \hat{F}(k;p)\right| \leq \sqrt{N_p}\,\sqrt{\sum_{\boldsymbol{\alpha}_p \in \Lambda} \left|F_k(\boldsymbol{\alpha}_p) - \hat{F}(k;p)\right|^2}$$

Besides the PSF approximation error, as another source of error for a diffractive function approximator, we also consider spatially varying, pixel-dependent, input phase error, $\Delta\phi_p(\mathbf{a})$, which can arise due to, e.g., illumination noise, SLM errors or even limited phase bit depth of the optical hardware. This can produce a field perturbation, creating an additional field-based function approximation error bound ($E_{\phi,k}$). Assume that the input phase due to this spatially varying phase error can be written as $2\pi\,\boldsymbol{\alpha}_p \cdot \mathbf{a} + \Delta\phi_p(\mathbf{a})$, which results in an inaccurate complex output field $\tilde{f}_k^{(\Delta\phi)}(\mathbf{a})$. Then, we have the following complex field error:

$$\tilde{f}_k^{(\Delta\phi)}(\mathbf{a}) - \tilde{f}_k(\mathbf{a}) = \sum_{p=1}^{N_p} \hat{F}(k;p) \exp\left(j2\pi\,\boldsymbol{\alpha}_p \cdot \mathbf{a}\right) \left[\exp\left(j\Delta\phi_p(\mathbf{a})\right) - 1\right]$$

This can be written as:

$$\left|\tilde{f}_k^{(\Delta\phi)}(\mathbf{a}) - \tilde{f}_k(\mathbf{a})\right| \leq \sum_{p=1}^{N_p} \left|\hat{F}(k;p)\right| \left|\exp\left(j\Delta\phi_p(\mathbf{a})\right) - 1\right| = 2\sum_{p=1}^{N_p} \left|\hat{F}(k;p)\right| \left|\sin\left(\frac{\Delta\phi_p(\mathbf{a})}{2}\right)\right| \quad \textbf{Eq. (6)}$$

We can also relate this analysis to Eq. (5), by writing:

$$f_{k,\Lambda}(\mathbf{a}) - \tilde{f}_k^{(\Delta\phi)}(\mathbf{a}) = \sum_{p=1}^{N_p} \left[c_{k,p} + \hat{F}(k;p)\left(1 - \exp\left(j\Delta\phi_p(\mathbf{a})\right)\right)\right] \exp\left(j2\pi\,\boldsymbol{\alpha}_p \cdot \mathbf{a}\right).$$

Taking the supremum over $K$, we can get a new inequality that combines $E_{\text{PSF},k}$ with $E_{\phi,k}$, given by **Eq. (7)**:

$$\| f_{k,\Lambda}(\mathbf{a}) - \tilde{f}_k^{(\Delta\phi)}(\mathbf{a}) \|_{\infty,K} \leq \sum_{p=1}^{N_p} \left|c_{k,p}\right| + \sup_{\mathbf{a}\in K} \sum_{p=1}^{N_p} \left|\hat{F}(k;p)\right| \left|1 - \exp\left(j\Delta\phi_p(\mathbf{a})\right)\right| = E_{\text{PSF},k} + E_{\phi,k}$$

where the input phase error induced function approximation error bound $E_{\phi,k}$ is given by:

$$E_{\phi,k} = \sup_{\mathbf{a}\in K} \sum_{p=1}^{N_p} \left|\hat{F}(k;p)\right| \left|1 - \exp\left(j\Delta\phi_p(\mathbf{a})\right)\right|$$

$$= 2 \sup_{\mathbf{a}\in K} \sum_{p=1}^{N_p} \left|\hat{F}(k;p)\right| \left|\sin\left(\frac{\Delta\phi_p(\mathbf{a})}{2}\right)\right| \qquad \textbf{Eq. (8A)}$$

For a bounded phase error, where $\left|\Delta\phi_p(\mathbf{a})\right| \leq \delta_\phi$, we can rewrite Eq. (8A) as:

$$E_{\phi,k} \leq \delta_\phi \sum_{p=1}^{N_p} \left|\hat{F}(k;p)\right| \quad \textbf{Eq. (8B)}$$

The combined error bound reported above in Eq. (7) is numerically validated by our analysis in Fig. 5. We tested two different diffractive processor configurations with (1) $N = 1.25 \times 2N_pN_f$, $N_f = 10{,}000$, $K_L = 2$ and (2) $N = 1.25 \times 2N_pN_f$, $N_f = 99{,}856$, $K_L = 4$. For both configurations, the input phase error, $\Delta\phi_p$, was simulated by applying 8-bit phase quantization to the desired phase values, $2\pi\boldsymbol{\alpha}_p \cdot \mathbf{a}$, during blind testing of the optical processors that were trained under the assumption of infinite phase precision. Our results in Fig. 5 reveal that all individual $\left(E_{\mathrm{PSF},k}, E_{\phi,k}, E_{f,k}\right)$ points corresponding to $k = 1, \dots, N_f$ nonlinear functions (implemented all in parallel) lie below the plane $E_f = E_{\mathrm{PSF}} + E_\phi$. This confirms that the combined PSF-synthesis and input phase-error terms provide a conservative bound on the resulting function-approximation maximum error, following Eq. (7).

Note that an analogous bound also holds for input amplitude modulation errors. If the encoded input field contains small field amplitude perturbations ($m_p$) that are spatially varying, represented by $1 + m_p(\mathbf{a})$, then we have:

$$E_{m,k} \le \delta_m \sum_{p=1}^{N_p} \left|\hat{F}(k;p)\right| \text{ for } \left|m_p(\mathbf{a})\right| \le \delta_m \text{ where } E_{m,k} = \sup_{\mathbf{a}\in K} \sum_{p=1}^{N_p} \left|\hat{F}(k;p)\right| \left|m_p(\mathbf{a})\right|.$$

In the case of small phase and amplitude modulation errors that exist together at the optical encoder, we will have $u_{\mathrm{in}}(p;\mathbf{a}) = \left[1 + m_p(\mathbf{a})\right] \exp\left(j2\pi\, \boldsymbol{\alpha}_p \cdot \mathbf{a} + j\Delta\phi_p(\mathbf{a})\right)$, where $\left|\Delta\phi_p(\mathbf{a})\right| \le \delta_\phi$ and $\left|m_p(\mathbf{a})\right| \le \delta_m$. For this condition, we can rewrite the equivalent of Eq. (7) as:

$$\| f_{k,\Lambda}(\mathbf{a}) - \tilde{f}_k^{(\Delta\phi,m)}(\mathbf{a}) \|_{\infty,K} \le E_{\mathrm{PSF},k} + \sqrt{\delta_m^2 + 2(1+\delta_m)(1-\cos\delta_\phi)}\ \sum_{p=1}^{N_p} \left|\hat{F}(k;p)\right|.$$

If we assume small angle and small amplitude errors, the above equation can be approximated as:

$$\| f_{k,\Lambda}(\mathbf{a}) - \tilde{f}_k^{(\Delta\phi,m)}(\mathbf{a}) \|_{\infty,K} \lesssim E_{\mathrm{PSF},k} + \sqrt{{\delta_m}^2 + {\delta_\phi}^2}\ \sum_{p=1}^{N_p} \left|\hat{F}(k;p)\right|.$$

For the sake of simplicity, going forward we assume negligible amplitude modulation error, i.e., $\left|m_p(\mathbf{a})\right| \approx 0$ and Eqs. (7) and (8A) both hold.

In our analysis, the full complex field-based function approximation error bound with respect to the true target function $f_k(\mathbf{a})$ should also be considered, which includes additional terms, i.e.,

$$\| f_k(\mathbf{a}) - \tilde{f}_k^{(\Delta\phi)}(\mathbf{a}) \|_{\infty,K} \le E_{\mathrm{PSF},k} + E_{\phi,k} + E_{\mathrm{trunc},k} \quad \textbf{Eq. (9A)}$$

where the Fourier truncation or bandlimiting error upper bound is defined as:

$$E_{\mathrm{trunc},k} = \| f_k(\mathbf{a}) - f_{k,\Lambda}(\mathbf{a}) \|_{\infty,K}$$
$$\le \sum_{\boldsymbol{\alpha}\notin\Lambda} \left|F_k(\boldsymbol{\alpha})\right| \quad \textbf{Eq. (9B)}$$

The inequality on the second line of Eq. (9B) assumes that the target function $f_k(\mathbf{a})$ has a convergent discrete Fourier-series representation on the compact domain of interest $K \subset \mathbb{R}^D$ and its Fourier coefficients are absolutely summable.

We also need to account for additional sources of error in the $\tilde{f}_k^{(\Delta\phi)} \rightarrow \hat{f}_k$ process for the measured or decoded complex field, i.e.,

$$\| f_k(\mathbf{a}) - \hat{f}_k(\mathbf{a}) \|_{\infty,K} \leq E_{\mathrm{PSF},k} + E_{\phi,k} + E_{\mathrm{trunc},k} + E_{\mathrm{misc},k} = \epsilon_k \quad \textbf{Eq. (10)}$$

Here, $E_{\mathrm{misc,k}}$ is a lumped term for additional/remaining nonidealities, which may include, e.g., optical hardware errors, readout/decoding errors, and residual imperfections not already included in the other error terms. Some implementation nonidealities can be treated after design as additive error terms, while others can be included directly in the optical forward model during the training of the diffractive processor. For example, absorption, surface reflections, aperture truncation, mechanical misalignments, and fabrication imperfections of a diffractive processor can be modeled during the optimization of the diffractive layers, partially reducing their contributions to the final $E_{\mathrm{misc,k}}$.

Earlier results[14–19] report that synthesizing an arbitrary complex-valued $N_f \times N_p$ transformation matrix using a diffractive optical processor requires at least $N \gtrsim 2N_pN_f$ independent controllable/optimizable phase degrees of freedom; note that this condition can be reduced to $N \gtrsim N_pN_f$ if the trainable diffractive features of an optical processor are modeled as complex-valued amplitude-and-phase elements. In our formalism in this paper, we consider phase-only diffractive features due to their ease of physical implementation, e.g., via SLMs and 3D nanofabrication methods[23–25]. The necessary condition for a phase-only diffractive processor, i.e., $N \gtrsim 2N_pN_f$, gives a parameter-count (phase degrees-of-freedom) heuristic for exact synthesis; earlier works have demonstrated task-specific approximation accuracy for smaller $N$ values, i.e., $N < 2N_pN_f$, using deeper diffractive architectures where the controllable phase degrees of freedom are spread over successive layers as opposed to being concentrated into a shallow diffractive architecture, which increases the approximation error due to contributions from ballistic photons[14–17]. On the other hand, prior work also used slightly overparameterized diffractive designs for achieving better local minima and reduced error in the optimization of a multiplexed diffractive function approximator; for example, $N \approx 1.25 \times 2N_pN_f$ phase features were used in some of our earlier designs[11]. However, the achieved function approximation error still depends on diffractive architecture/geometry, aperture size, layer count, light throughput constraints, optimization landscape and tolerances of the physical set-up. Increasing the number ($K_L$) of diffractive layers does not by itself increase the input/output space-bandwidth product, but it does improve controllability, PSF synthesis accuracy, ballistic-light suppression, and output diffraction efficiency[11].

For a general $D$-variate nonlinear function expanded on a tensor-product Fourier grid with $F$ frequency samples along each dimension, the number of required Fourier features scales as $N_p = F^D$ which indicates an exponential growth in the number of diffractive phase features required, i.e., $N \gtrsim 2F^D N_f$. For the general case of nonuniform frequency samples along different dimensions, the same expression can be written as $N_p = \prod_{d=1}^{D} F_d$, i.e.,

$$N \gtrsim 2N_f \prod_{d=1}^{D} F_d \quad \textbf{Eq. (11)}$$

where, $F_d$ is the number of selected frequency samples along dimension $d$. This indicates that, although diffractive optical processors provide massive physical parallelism with a large $N_f$ and a large space-bandwidth product, their architecture does not remove the curse of dimensionality in approximating high-dimensional nonlinear functions with $D \gg 1$ and $F \gg 1$, where a massive increase in $N \gtrsim 2F^D N_f$ would be required. Therefore, while an optical diffractive processor reduces latency and can exploit very large numbers of spatial modes, the number of basis functions ($N_p$) needed at the input illumination aperture of the diffractive processor to accurately approximate an arbitrary high-dimensional function remains governed by the structure of the target function class, with a scaling law dictated by $N_p = F^D$ or $N_p = \prod_{d=1}^{D} F_d$. More favorable scaling is possible for structured target functions, including functions with sparse spectra, low effective dimensions or low-rank tensor structures.

To demonstrate this framework in a representative high-dimensional function approximation setting, we trained a diffractive optical processor (with $K_L = 8$, $N = 50\, F^D N_f$ and $D = F = 4$) to approximate $N_f = 4$ independent 4-D target nonlinear functions that are complex-valued, i.e., $f_1(a_1, a_2, a_3, a_4)$, $f_2(a_1, a_2, a_3, a_4)$, $f_3(a_1, a_2, a_3, a_4)$, and $f_4(a_1, a_2, a_3, a_4)$. The corresponding optical architecture is shown schematically in Fig. 6, while the learned 4-D nonlinear function approximation results are shown in Figs. 7-8, all of which confirm low MSE values with respect to the ground truth nonlinear functions $f_1$, $f_2$, $f_3$, and $f_4$. These results demonstrate that the diffractive function approximator can accurately approximate multiple complex-valued high-dimensional target nonlinear functions when the required finite Fourier-feature representation is physically encoded at the input aperture and synthesized through the trained, task-optimized diffractive layers.

The above analyses focused on the complex field-based function approximation error of a diffractive optical processor; see Eqs. (5-10). For intensity-only readout at the output plane, a field-based approximation maximum error bound of $\| f_k(\mathbf{a}) - \hat{f}_k(\mathbf{a}) \|_{\infty,K} \leq \epsilon_k$ given by Eq. (10) induces an intensity-based function approximation maximum error bound, defined as $\| \, |f_k(\mathbf{a})|^2 - \left|\hat{f}_k(\mathbf{a})\right|^2 \|_{\infty,K}$. This intensity-based error bound applies when the target function is represented with $g_k(\mathbf{a}) = |f_k(\mathbf{a})|^2$. To expand on this intensity-based function approximation error, for each input variable $\mathbf{a} = (a_1, \ldots, a_D)$ over the compact domain of interest, $K \subset \mathbb{R}^D$, we can write:

$$\left| |f_k(\mathbf{a})|^2 - \left|\hat{f}_k(\mathbf{a})\right|^2 \right| = \left| |f_k(\mathbf{a})| - \left|\hat{f}_k(\mathbf{a})\right| \right| \left( |f_k(\mathbf{a})| + \left|\hat{f}_k(\mathbf{a})\right| \right).$$

By the reverse triangle inequality, we have

$$\left| |f_k(\mathbf{a})| - \left|\hat{f}_k(\mathbf{a})\right| \right| \leq \left| f_k(\mathbf{a}) - \hat{f}_k(\mathbf{a}) \right|.$$

Therefore, we get,

$$\left||f_k(\mathbf{a})|^2 - \left|\hat{f}_k(\mathbf{a})\right|^2\right| \leq \left|f_k(\mathbf{a}) - \hat{f}_k(\mathbf{a})\right| \left(|f_k(\mathbf{a})| + \left|\hat{f}_k(\mathbf{a})\right|\right)$$

To further simplify this term, we then use the following inequalities:

$$\left|f_k(\mathbf{a}) - \hat{f}_k(\mathbf{a})\right| \leq \| f_k(\mathbf{a}) - \hat{f}_k(\mathbf{a}) \|_{\infty,K} \leq \epsilon_k$$
$$|f_k(\mathbf{a})| \leq \| f_k(\mathbf{a}) \|_{\infty,K} \;\; and \;\; \left|\hat{f}_k(\mathbf{a})\right| \leq \| \hat{f}_k(\mathbf{a}) \|_{\infty,K}$$

which help us write:

$$\left||f_k(\mathbf{a})|^2 - \left|\hat{f}_k(\mathbf{a})\right|^2\right| \leq \| f_k(\mathbf{a}) - \hat{f}_k(\mathbf{a}) \|_{\infty,K} \left(\| f_k(\mathbf{a}) \|_{\infty,K} + \| \hat{f}_k(\mathbf{a}) \|_{\infty,K}\right)$$

Taking the supremum over $\mathbf{a} \in K$ gives:

$$\| \, |f_k(\mathbf{a})|^2 - \left|\hat{f}_k(\mathbf{a})\right|^2 \|_{\infty,K} \leq \| f_k(\mathbf{a}) - \hat{f}_k(\mathbf{a}) \|_{\infty,K} \left(\| f_k(\mathbf{a}) \|_{\infty,K} + \| \hat{f}_k(\mathbf{a}) \|_{\infty,K}\right)$$

Finally, since we have $\| f_k(\mathbf{a}) - \hat{f}_k(\mathbf{a}) \|_{\infty,K} \leq \epsilon_k$ and $\| \hat{f}_k(\mathbf{a}) \|_{\infty,K} \leq \| f_k(\mathbf{a}) \|_{\infty,K} + \epsilon_k$, the intensity-based function approximation maximum error bound of a diffractive optical processor can be written as:

$$\begin{aligned} \| \, |f_k(\mathbf{a})|^2 - \left|\hat{f}_k(\mathbf{a})\right|^2 \|_{\infty,K} &\leq \epsilon_k \left(\| f_k(\mathbf{a}) \|_{\infty,K} + \| \hat{f}_k(\mathbf{a}) \|_{\infty,K}\right) \\ &\leq \epsilon_k \left(2 \| f_k(\mathbf{a}) \|_{\infty,K} + \epsilon_k\right) \end{aligned}$$

which yields:

$$\| \, |f_k(\mathbf{a})|^2 - \left|\hat{f}_k(\mathbf{a})\right|^2 \|_{\infty,K} \leq 2\epsilon_k \| f_k(\mathbf{a}) \|_{\infty,K} + \epsilon_k^2 \qquad \mathbf{Eq.\,(12)}$$

More generally, if a complex-valued target nonlinear function, $g_k(\mathbf{a})$, is to be approximated using multiple nonnegative intensity channels at the output plane of a diffractive processor, we can express the target function as a linear combination of intensity values, i.e., $g_k(\mathbf{a}) = \sum_{x=0}^{X-1} w_x \left|f_{k,x}(\mathbf{a})\right|^2$, where $w_x$ are complex weights. Using intensity readout at the output of a diffractive optical processor, we can get: $\hat{g}_k(\mathbf{a}) = \sum_{x=0}^{X-1} w_x \left|\hat{f}_{k,x}(\mathbf{a})\right|^2$. Based on this multi-basis intensity measurement scheme, the total complex-valued nonlinear function approximation error will be bounded by the weighted sum of the individual intensity channel errors, i.e.,

$$\| g_k(\mathbf{a}) - \hat{g}_k(\mathbf{a}) \|_{\infty,K} \leq \sum_{x=0}^{X-1} |w_x| \; \| \left|f_{k,x}(\mathbf{a})\right|^2 - \left|\hat{f}_{k,x}(\mathbf{a})\right|^2 \|_{\infty,K}$$

Applying Eq. (12) to each intensity channel gives the total multi-basis intensity-readout error bound, i.e.,

$$\| g_k(\mathbf{a}) - \hat{g}_k(\mathbf{a}) \|_{\infty,K} \leq \sum_{x=0}^{X-1} |w_x| \left[2\epsilon_{k,x} \| f_{k,x}(\mathbf{a}) \|_{\infty,K} + \epsilon_{k,x}^2\right] \quad \textbf{Eq. (13)}$$

As an example, consider $g_k(\mathbf{a}) = \sum_{x=0}^{2} w_x \left|f_{k,x}(\mathbf{a})\right|^2$ with $w_x = \exp\left(jx\frac{2\pi}{3}\right)$, which can be used to efficiently approximate a bounded complex-valued nonlinear function using $X = 3$ intensity readout values at the output of a diffractive processor[26,27], and therefore the total intensity readout error in this case is bounded by the sum of the three intensity channel error bounds, following Eq. (13) with $|w_x| = 1$.

**Energy–Throughput–Precision Tradeoffs**

Passive diffractive layers do not consume active energy while the optical field propagates through them. The full system, however, still consumes energy for illumination, input modulation, detector/readout electronics, and control. Assume the intensity readout of a diffractive optical processor under ideal Poisson shot noise is characterized by,

$$C_k \sim \text{Poisson}(N_{\text{det},k}),$$

where $\mathbb{E}[C_k] = \text{Var}(C_k) = N_{\text{det},k}$ and $C_k$ is the detected photon count in the output channel $k$ of a diffractive function approximator. Hence, the standard deviation of the photon count is $\sigma_{C_k} = \sqrt{N_{\text{det},k}}$. Therefore, the relative shot-noise fluctuation can be written as:

$$\frac{\sigma_{C_k}}{\mathbb{E}[C_k]} = \frac{1}{\sqrt{N_{\text{det},k}}}.$$

For a normalized intensity output, the photon-count fluctuation therefore induces a normalized uncertainty on the order of $\frac{1}{\sqrt{N_{\text{det},k}}}$. To resolve the normalized output intensity at the k-th detector (corresponding to the function $\left|\tilde{f}_k(\mathbf{a})\right|^2$) with a relative precision of $\varepsilon_k$, the shot-noise uncertainty must satisfy $\frac{1}{\sqrt{N_{\text{det},k}}} \lesssim \varepsilon_k$. Therefore, for the output channel $k$ of a diffractive function approximator, resolving a normalized output intensity with a relative precision $\varepsilon_k$ requires a detected photon number on the order of:

$$N_{\text{det},k} \gtrsim \frac{1}{\varepsilon_k^2}$$

If $D_k(\mathbf{a})$ is the diffraction efficiency of the diffractive function approximator into the relevant output aperture (i.e., the detector active region) of the k-th target function ($|f_k(\mathbf{a})|^2 \approx \left|\tilde{f}_k(\mathbf{a})\right|^2$), we have:

$$N_{\text{det},k} = \eta_k\, D_k(\mathbf{a})\, N_{\text{in}} \gtrsim \frac{1}{\varepsilon_k^2}$$

where $\eta_k$ is the external quantum efficiency of the output detector $k$, and $N_{\text{in}}$ is the incident photon number in the optical inference of $N_f$ parallel output functions. To achieve a desired relative precision $\varepsilon_k$ at every required output channel $k$ under a shared illumination budget, we need to achieve:

$$N_{\text{in}} \gtrsim \max_k \frac{1}{\eta_k\, D_{k,min}\, \varepsilon_k^2}.$$

where $D_{k,\text{min}} = \inf_{\mathbf{a}\in K} D_k(\mathbf{a})$. This results in the corresponding shot-noise-limited input energy estimate:

$$E_{\text{in}} \gtrsim \frac{hc}{\lambda} \max_k \frac{1}{\eta_k\, D_{k,\text{min}}\, \varepsilon_k^2} \quad \textbf{Eq. (14)}$$

where $\lambda$ is the illumination wavelength, $h$ is Planck's constant, and $c$ is the speed of light. If the diffraction efficiencies are design variables constrained by a total throughput budget, for example $\sum_{k=1}^{N_f} D_k(\mathbf{a}) = D_{\text{total}}(\mathbf{a})$, then uniform precision across many output channels can still introduce scaling with the number of readout channels ($k = 1,2,\dots,N_f$) through the appropriate allocation of the input optical power. For example, in the uniform energy allocation case with balanced channel efficiencies, if $\eta_k = \eta$, $\varepsilon_k = \varepsilon$, and $D_{k,\text{min}} \approx D_{\text{total,min}}/N_f$, we have:

$$N_{\text{in}} \gtrsim \frac{N_f}{\eta D_{\text{total,min}} \varepsilon^2}, \quad E_{\text{in}} \gtrsim \frac{hc}{\lambda} \frac{N_f}{\eta D_{\text{total,min}} \varepsilon^2} \quad \textbf{Eq. (15)}$$

Here, $D_{\text{total,min}} = \inf_{\mathbf{a}\in K} D_{\text{total}}(\mathbf{a})$. Equation (15) indicates that the cost of parallelism in a diffractive optical processor with a large $N_f$ is the photon budget required to simultaneously resolve many output channels.

This analysis reflects a photon-statistics- and throughput-bound under idealized detection assumptions. This relative-precision estimate applies to output channels with non-negligible detected intensity. For target detectors that approach zero signal, an absolute-error photon budget or a differential/baseline readout model should be used. Furthermore, the above estimate treats the normalization or calibration reference as fixed. If the normalization denominator is itself measured optically, its shot noise and detector noise must also be propagated into the final uncertainty.

Through empirical analysis via deep learning-based optimization of diffractive function approximators, our former work[11] showed an important practical trade-off: constraints that improve the output diffraction efficiency, $D_{\text{total}}$, can slightly degrade function-approximation accuracy, while deeper diffractive architectures can reduce ballistic photon leakage, and improve both accuracy and total diffraction efficiency for various optical processor designs. The same conclusions on the advantages of deeper diffractive processor designs have also been validated for other computational tasks such as optical phase conjugation, quantitative phase imaging, image classification or linear transformations[14,15,28–31].

The above analyses were specific to intensity-based detection at the output of a diffractive processor. We can also perform a related analysis for complex field-based nonlinear function approximation through a diffractive optical processor. For shot noise-like finite-photon phase-encoding noise $\Delta\phi_p(\mathbf{a})$, we can assume that $\mathrm{Var}\left(\Delta\phi_p(\mathbf{a})\right) = \frac{\gamma}{n_p} \approx O\left(\frac{1}{n_p}\right)$ and $\sigma\left(\Delta\phi_p(\mathbf{a})\right) \approx O\left(\frac{1}{\sqrt{n_p}}\right)$, where $\gamma$ is a readout/measurement set-up specific constant and $n_p > 0$ is the mean photon number at the input encoder pixel $p$, i.e., $n_p = \beta_p N_{\text{in}}$ with $\sum_{p=1}^{N_p} \beta_p = 1$. Following the same notation as in Eqs. (6-8), and assuming small $\Delta\phi_p(\mathbf{a})$ fluctuations, we have:

$$\tilde{f}_k^{(\Delta\phi)}(\mathbf{a}) - \tilde{f}_k(\mathbf{a}) = \sum_{p=1}^{N_p} \hat{F}(k;p) \exp\left(j2\pi\, \boldsymbol{\alpha}_p \cdot \mathbf{a}\right) \left[\exp\left(j\Delta\phi_p(\mathbf{a})\right) - 1\right]$$
$$\approx j \sum_{p=1}^{N_p} \Delta\phi_p(\mathbf{a})\, \hat{F}(k;p) \exp\left(j2\pi\, \boldsymbol{\alpha}_p \cdot \mathbf{a}\right)$$

Then we can write:

$$\mathbb{E}\left[\left|\tilde{f}_k^{(\Delta\phi)}(\mathbf{a}) - \tilde{f}_k(\mathbf{a})\right|^2\right]$$
$$= \mathbb{E}\left[\sum_{p=1}^{N_p} \sum_{p'=1}^{N_p} \Delta\phi_p(\mathbf{a})\, \Delta\phi_{p'}(\mathbf{a}) \hat{F}(k;p) \hat{F}^*(k;p') \exp\left(j2\pi \left[\boldsymbol{\alpha}_p - \boldsymbol{\alpha}_{p'}\right] \cdot \mathbf{a}\right)\right]$$

Assuming that the phase fluctuations are independent with zero-mean, for $p \neq p'$ we have $\mathbb{E}\left[\Delta\phi_p(\mathbf{a})\, \Delta\phi_{p'}(\mathbf{a})\right] = 0$, which can be used to simplify the above equation as:

$$\mathbb{E}\left[\left|\tilde{f}_k^{(\Delta\phi)}(\mathbf{a}) - \tilde{f}_k(\mathbf{a})\right|^2\right] = \mathbb{E}\left[\sum_{p=1}^{N_p} \Delta\phi_p(\mathbf{a})^2 \left|\hat{F}(k;p)\right|^2\right] = \sum_{p=1}^{N_p} \frac{\gamma}{n_p} \left|\hat{F}(k;p)\right|^2$$

$$= \sum_{p=1}^{N_p} \frac{\gamma}{\beta_p N_{\text{in}}} \left|\hat{F}(k;p)\right|^2 \quad \textbf{Eq. (16)}$$

To further simplify this equation, we can assume equal distribution of input photons at each phase encoder pixel, i.e., $N_{\text{in}} = N_p n_p$ with a fixed $\beta_p = 1/N_p$ to arrive at the following $\Delta\phi_p(\mathbf{a})$ induced complex field-based function approximation mean squared error:

$$\mathbb{E}\left[\left|\tilde{f}_k^{(\Delta\phi)}(\mathbf{a}) - \tilde{f}_k(\mathbf{a})\right|^2\right] = \frac{\gamma N_p}{N_{\text{in}}} \sum_{p=1}^{N_p} \left|\hat{F}(k;p)\right|^2 \approx O\left(\frac{N_p}{N_{\text{in}}} \sum_{p=1}^{N_p} \left|\hat{F}(k;p)\right|^2\right) \quad \textbf{Eq. (17)}$$

As stated earlier, this estimate assumes independent, zero-mean, small phase fluctuations. Also note that Eqs. (16-17) are probabilistic estimates of $\Delta\phi_p(\mathbf{a})$ induced complex field-based function approximation mean-squared error; for deterministic uniform bounds over $\mathbf{a} \in K$, the readers should refer to Eqs. (7-8).

**Statistical Learnability and Training Sample Complexity for Diffractive Function Approximators**

The universal function approximation theorem and statistical learning address different questions[1,10]. A universal approximation result indicates that, under the stated assumptions, some diffractive designs can approximate a desired function or family of functions. A sample-complexity result instead asks how many sampled training pairs are statistically needed so that a trained diffractive function approximator that performs well on the training samples is also expected to perform well on new, unseen inputs – achieving generalization. The universal function approximation statement concerns representability; the sample-complexity statement explored in this sub-section concerns generalization from finite training data.

One way to write the diffractive hypothesis class is:

$$\mathcal{H}_{\text{diff}} = \left\{h_{\boldsymbol{\phi}}: \mathbf{a} \mapsto \mathcal{R}\left(T_{\boldsymbol{\phi}} u_{\text{in}}(\mathbf{a})\right): \boldsymbol{\phi} \text{ ranges over } \Phi\right\}.$$

Here $\mathcal{H}_{\text{diff}}$ is the set of all functions that the diffractive optical system can realize as its trainable optical parameters vary. Same as in the earlier sub-sections, $\mathbf{a}$ denotes the input variable of the target nonlinear function: for a $D$-dimensional complex-valued function $\mathbf{a} = (a_1, \ldots, a_D) \in \mathbb{R}^D$. The function $h_{\boldsymbol{\phi}}(\mathbf{a})$ is the output produced by the diffractive processor when its trainable phase pattern is $\boldsymbol{\phi}$. The input optical field $u_{\text{in}}(\mathbf{a})$ is the wavefront created by encoding $\mathbf{a}$ across the input pixels, same as described in earlier sections. The operator $T_{\boldsymbol{\phi}}$ is the complex-valued optical transformation operator produced by the diffractive processor for a particular trainable phase vector $\boldsymbol{\phi}$. The vector $\boldsymbol{\phi} = (\phi_1, \ldots, \phi_N)$ contains the $N$ trainable phase parameters, where $\phi_i$ is the phase delay imposed by the $i$-th diffractive feature, spatial-light-modulator pixel, or fabricated phase element (depending on the physical hardware of the diffractive optical processor design). $\mathcal{R}$ converts the output optical field $T_{\boldsymbol{\phi}} u_{\text{in}}(\mathbf{a})$ into a numerical function value. It may represent coherent complex-field readout, direct intensity detection, normalized intensity, differential detection, or separate real and imaginary channels, as demonstrated in our former work.[11]

The admissible phase space $\Phi$ is the set of allowed phase patterns. For continuous phase modulation, we can write it as:

$$\Phi = \{\boldsymbol{\phi}: 0 \leq \phi_i < 2\pi \text{ for each } i = 1, \ldots, N\}.$$

However, an infinite phase bit-depth is not practical for a given diffractive processor hardware, and therefore, we introduce a finite phase bit-depth of $B$ for each diffractive feature, $\phi_i$. In this case, each phase parameter $\phi_i$ can take only $2^B$ discrete values. Therefore, the total number of possible phase structures of a diffractive function approximator is at most $2^{BN}$. If each state of the diffractive processor defines at most one realized function, then the number of hypotheses is also bounded by $|\mathcal{H}_{\text{diff}}| \leq 2^{BN}$. This count assumes that the input encoding, frequency set, detector layout, normalization rule, and readout map $\mathcal{R}$ are fixed. The number of input pixels or Fourier features $N_p$, the number of output detector channels or target functions $N_f$, the number of diffractive layers ($K_L$), input/output numerical aperture, wavelength, detector layout, and noise level all influence physically realizable forms of $T_{\boldsymbol{\phi}}$ and $\mathcal{R}$ functions.

For simplicity, in our initial analysis, we consider one output channel with complex-field readout, i.e., $h_{\boldsymbol{\phi}}(\mathbf{a}) = \tilde{f}_k(\mathbf{a})$, where the k-th output channel of a diffractive function approximator is considered for a complex-valued, cascadable function approximation using the output complex field, $\tilde{f}_k(\mathbf{a})$.

Our training data set, with $M$ denoting the number of training samples, can be written as:

$$S_M = \{(\mathbf{a}_i, y_i)\}_{i=1}^{M}.$$

The target value $y_i = f_k(\mathbf{a}_i)$ is the desired output at input $\mathbf{a}_i$. At this point, it is useful to distinguish two learning problems. The first is the statistical learning of the target nonlinear function from samples $(\mathbf{a}_i, f_k(\mathbf{a}_i))$. The second is physical training of the diffractive hardware to synthesize a desired PSF matrix or input-output transformation. Deep-learning-based diffractive processor designs address the second problem by training diffractive layers using input-output field pairs associated with a target linear transformation that represents a desired spatially varying PSF set[14]. The finite-class bound that we analyze below instead controls the empirical-to-expected loss gap for a quantized diffractive hypothesis class for nonlinear function approximation.

Assume we have a loss function $\ell_k\left(\tilde{f}_k(\mathbf{a}_i), f_k(\mathbf{a}_i)\right)$ that measures the training penalty for each sample, which can be expressed over the entire training set of $M$ samples as:

$$\hat{L}_{M,k}(\boldsymbol{\phi}) = \frac{1}{M}\sum_{i=1}^{M} \ell_k\left(\tilde{f}_k(\mathbf{a}_i), f_k(\mathbf{a}_i)\right).$$

The expected/true loss is:

$$L_k(\boldsymbol{\phi}) = \mathbb{E}_{(\mathbf{a}_i, y_i)\sim P}\left[\ell_k\left(\tilde{f}_k(\mathbf{a}), f_k(\mathbf{a})\right)\right],$$

where $P$ is the probability distribution that generates the inputs and targets; for simplicity, we assume the loss function is real-valued. Generalization of the diffractive function approximator means that the empirical loss $\hat{L}_{M,k}$ is close to the true loss $L_k$, not merely that the diffractive optical system with $\boldsymbol{\phi}$ phase features fits the sampled training points. A typical high-probability generalization performance would indicate that, with a probability of at least $1-\delta$ over the random draw of the $M$ training samples, the gap between the true and empirical loss is at most $E$. The parameter $E$ is the tolerated generalization gap, and $\delta$ is the maximum allowed failure probability (Pr), i.e., the maximum probability that the generalization bound fails over the random draw of the training set.

For a given fixed $\boldsymbol{\phi}$ diffractive design ($h_{\boldsymbol{\phi}} \in \mathcal{H}_{\text{diff}}$), and with a bounded loss in $[a, b]$, i.e., $a \leq \ell_k\left(\tilde{f}_k(\mathbf{a}_i), f_k(\mathbf{a}_i)\right) \leq b$, Hoeffding's inequality[1,10] controls the probability that $L_k(\boldsymbol{\phi})$ and $\hat{L}_{M,k}(\boldsymbol{\phi})$ differ by at least $E$ , i.e.:

$$\Pr\left\{\left|\hat{L}_{M,k}(\boldsymbol{\phi}) - L_k(\boldsymbol{\phi})\right| \geq E\right\} \leq 2\exp(-2\text{M}E^2 / (\text{b} - a)^2) \quad \textbf{Eq. (18)}$$

For a diffractive function approximator with a phase bit-depth of $B$, and $N$ trainable diffractive features, there are at most $2^{BN}$ hypotheses in $\mathcal{H}_{\text{diff}}$ , and therefore we can apply a union bound over all possible hypotheses to arrive at:

$$\Pr\left\{\sup_{h_{\boldsymbol{\phi}} \in \mathcal{H}_{\text{diff}}} \left|\hat{L}_{M,k}(\boldsymbol{\phi}) - L_k(\boldsymbol{\phi})\right| \geq E\right\} \leq 2^{BN}\, 2\exp(-2\text{M}E^2 / (\text{b} - a)^2) \quad \textbf{Eq. (19)}$$

Since the maximum allowed failure probability is $\delta$, we have:

$$2^{BN+1} \exp(-2\text{M}E^2 / (\text{b} - a)^2) \leq \delta \quad \textbf{Eq. (20)}$$

Eq. (20) can be expressed as:

$$\text{M} \geq (\text{b} - a)^2 \frac{BNln2 + \ln(\frac{2}{\delta})}{2E^2} \quad \textbf{Eq. (21)}$$

Therefore, the sample-complexity expression scales on the order ($O$) of:

$$\text{M} = O\left(\frac{BN + \ln(\frac{1}{\delta})}{E^2}\right) \quad \textbf{Eq. (22)}$$

This generalization condition implies that as the tolerated generalization gap $E$ decreases, the number of training samples (M) must scale as $1/E^2$. This gives a tighter uniform bound between the empirical and expected/true loss; it does not, by itself, guarantee a smaller total function-approximation error unless the empirical loss, optimization error, approximation error, and physical noise are also controlled. Similarly, as the maximum failure probability $\delta$ approaches ~0%, a substantial increase in M is required. However, the logarithmic dependence on $1/\delta$ means that increasing the confidence requirement (smaller $\delta$) increases the training sample size more mildly.

If the loss is a single aggregate loss ($\ell_{\text{agg}}$) over *all* $N_f$ output functions ($k = 1,2,\dots,N_f$), i.e., $\ell_{\text{agg}} = \frac{1}{N_f}\sum_{k=1}^{N_f} \ell_k$, then we have:

$$\hat{L}_{M,agg}(\boldsymbol{\phi}) = \frac{1}{M}\sum_{i=1}^{M} \ell_{\text{agg}}(\boldsymbol{h}_{\boldsymbol{\phi}}(\mathbf{a}_i), \mathbf{y}_i)$$

$$L_{agg}(\boldsymbol{\phi}) = \mathbb{E}_{(\mathbf{a}_i, \mathbf{y}_i)\sim P}\left[\ell_{\text{agg}}(\boldsymbol{h}_{\boldsymbol{\phi}}(\mathbf{a}_i), \mathbf{y}_i)\right],$$

where each training input $\mathbf{a}_i$ is paired with the vector of target values $\mathbf{y}_i = \left[f_1(\mathbf{a}_i), \dots, f_{N_f}(\mathbf{a}_i)\right]$, i.e., $S_M = \{(\mathbf{a}_i, \boldsymbol{y}_i)\}_{i=1}^{M}$ and $\boldsymbol{h}_{\boldsymbol{\phi}}(\mathbf{a}_i) = \left[\tilde{f}_1(\mathbf{a}_i), \dots, \tilde{f}_{N_f}(\mathbf{a}_i)\right]$. If we assume $a \le \ell_k\left(\tilde{f}_k(\mathbf{a}_i), f_k(\mathbf{a}_i)\right) \le b$ for all $k = 1,2, \dots, N_f$, then $a \le \ell_{\text{agg}} \le b$ also holds in this case, and we have:

$$\Pr\left\{\left|\hat{L}_{M,agg}(\boldsymbol{\phi}) - L_{agg}(\boldsymbol{\phi})\right| \ge E\right\} \le 2\exp(-2\mathrm{M}E^2 / (\mathrm{b}-a)^2)$$

Therefore, Eqs. (19-22) also hold here for the aggregate loss, i.e., $\mathrm{M} = O\left(\frac{BN + \ln(\frac{1}{\delta})}{E^2}\right)$, same as in Eq. (22).

The above analysis controls the aggregate empirical-to-expected loss gap, but it does not guarantee a separate bound for every output function channel. However, instead of this aggregate loss-based guarantee, if we seek a simultaneous per-output uniform guarantee for all $N_f$ function channels, then we should satisfy:

$$\sup_{h_{\boldsymbol{\phi}} \in \mathcal{H}_{\text{diff}}} \max_{1 \le k \le N_f} \left|\hat{L}_{M,k}(\boldsymbol{\phi}) - L_k(\boldsymbol{\phi})\right| \le E.$$

For this, we need an additional union bound over $N_f$ output channels, i.e., $2^{BN} \to N_f 2^{BN}$ in Eq. (19), which yields:

$$\Pr\left[\sup_{h_{\boldsymbol{\phi}} \in \mathcal{H}_{\text{diff}}} \max_{1 \le k \le N_f} \left|\hat{L}_{M,k}(\boldsymbol{\phi}) - L_k(\boldsymbol{\phi})\right| \ge E\right] \le 2\, N_f 2^{BN} \exp(-2\mathrm{M}E^2 / (\mathrm{b}-a)^2) \quad \textbf{Eq. (23)}$$

This Eq. (23) is similar to Eq. (19), except that it is for the simultaneous per-output uniform guarantee for all $N_f$ function channels. Following the same analysis as in Eqs. (19-22), the new Eq. (23) yields:

$$\mathrm{M} \ge (\mathrm{b}-a)^2 \, \frac{BNln2 + \ln(\frac{2N_f}{\delta})}{2E^2}$$

Using Eq. (11), we can assume $N \approx 2N_f \prod_{d=1}^{D} F_d \approx 2N_f F^D$ with a constant $F$ frequency samples along each dimension, which can be used to expand the above inequality as:

$$\mathrm{M} \ge (\mathrm{b}-a)^2 \, \frac{2BN_f F^D ln2 + \ln(\frac{2N_f}{\delta})}{2E^2} = (\mathrm{b}-a)^2 \, \frac{2BN_f N_p ln2 + \ln(\frac{2N_f}{\delta})}{2E^2}$$

$$\mathrm{M} = O\left(\frac{BN_f F^D + \ln(\frac{N_f}{\delta})}{E^2}\right) = O\left(\frac{BN_f N_p + \ln(\frac{N_f}{\delta})}{E^2}\right) \quad \textbf{Eq. (24)}$$

Therefore, in Eq. (24) the $\ln(N_f)$ term is due to the additional simultaneous-confidence penalty; however, the dominant $N_f$ dependence in Eq. (24) is due to the scaling of the total number of trainable physical/phase degrees of freedom, $N \approx 2N_f \prod_{d=1}^{D} F_d \approx 2N_f F^D$.

This finite-class analysis on the number of training samples, M, is useful as an intuition; it formulates an upper bound for a quantized diffractive model with a phase bit-depth of $B$. It indicates that increasing the number of trainable optical phase parameters $N$ or increasing their phase bit depth $B$ increases the number of possible phase masks and therefore requires an appropriate increase in the number of training samples needed for a uniform generalization guarantee. This bound assumes a finite quantized hypothesis class, fixed diffractive architecture, i.i.d. training samples, a bounded loss, and a learning algorithm whose selected design belongs to $\mathcal{H}_{\text{diff}}$. Using statistical learning theory[1,10], it bounds the empirical-to-expected loss gap, not the approximation error, optimization error, or physical implementation error. Stated differently, this finite-class bound concerns generalization from sampled input-output pairs and it does not bound the numerical optimization error of the diffractive layers, nor the physical PSF-synthesis error. Finally, we should note that these finite-class related analyses form a conservative worst-case capacity bound. Practical sample complexity may be smaller because many different states of diffractive phase layers might realize closely related optical transformations, or the learned nonlinear functions may have inherent low complexity.

**Coherent Cascadability and Depth of a Diffractive Processor**

As discussed earlier, a diffractive optical processor can output a coherent complex field, encoding a set of nonlinear functions, and a second diffractive processor can operate directly on that field without square-law detection[11]. This indicates complex field-level cascadability of the diffractive function approximator and can avoid intermediate optoelectronic conversion in tasks that preserve coherence.

However, this should not be described as a guarantee of exponential representation efficiency[32]. Traditional depth-separation results[32] concern nonlinear digital neural-network architectures. A cascade of passive linear diffractive systems remains linear with respect to the optical field entering the cascade. To further expand on this, let us consider the following complex optical field cascadable diffractive processor, where the first function approximator diffractive processor achieves various nonlinear functions $\tilde{f}_k(\mathbf{a}) = \sum_{\boldsymbol{\alpha}_p \in \Lambda} \hat{F}(k;p) \exp\left(j2\pi\, \boldsymbol{\alpha}_p \cdot \mathbf{a}\right)$, i.e., $f_k(\mathbf{a}) \approx \tilde{f}_k(\mathbf{a})$ for $k = 1,2,\dots,N_f$ and $\Lambda = \{\boldsymbol{\alpha}_p\}_{p=1}^{N_p}$; the second, cascaded diffractive processor then acts on these complex fields represented by $\tilde{f}_k(\mathbf{a})$ to create (at its own output plane) a complex field given by: $\sum_{k=1}^{N_f} \tilde{f}_k(\mathbf{a})\, \hat{G}(n;k)$, where $\hat{G}(n;k)$ represents the spatially varying complex-valued PSFs of the second diffractive processor, all-optically connecting the plane of $\tilde{f}_k(\mathbf{a})$ to the output plane of the 2nd diffractive processor, indexed by $n$.

Let's also assume that at the input plane of the second, cascaded diffractive processor, we allocate $N_p$, additional input channels, e.g., through an additional SLM that is connected to the same coherent input illumination source. These additional input channels are co-located with $\tilde{f}_k(\mathbf{a})$ and can be represented at the 2nd SLM plane as $i_{p'}(\mathbf{a}) = \mathrm{A}_{p'} \exp\left(j2\pi\, \boldsymbol{\alpha}_{p'} \cdot \mathbf{a}\right)$ where $\Lambda' = \{\boldsymbol{\alpha}_{p'}\}_{p'=1}^{N_{p'}}$ represents the finite frequency set implemented by these input pixels, and $0 \leq \mathrm{A}_{p'} \leq 1$ is a real-valued constant for each input channel $p'$. These additional channels $i_{p'}(\mathbf{a})$ require encoding $\mathbf{a}$ again at the second input plane, which can be achieved through, e.g., an additional SLM connected to the same illumination via beam splitters.

The complex output field at the output plane of the second diffractive processor can be written as a superposition of the contributions from $\tilde{f}_k(\mathbf{a})$ and $i_{p'}(\mathbf{a})$, i.e.,

$$\tilde{g}_n(\mathbf{a}) = \sum_{k=1}^{N_f} \tilde{f}_k(\mathbf{a})\, \hat{G}(n;k) + \sum_{p'=1}^{N_{p'}} i_{p'}(\mathbf{a})\, \widehat{G'}(n;p') \quad \textbf{Eq. (25)}$$

where $\hat{G}(n;k)$ and $\widehat{G'}(n;p')$ refer to the set of spatially varying PSFs that are jointly implemented by the 2nd cascaded diffractive optical processor, all-optically connecting the plane that is jointly indexed by $k$ and $p'$ to the output plane of the 2nd diffractive processor indexed by $n$.

Intensity detection at the output plane of the 2nd diffractive processor yields:

$$\begin{aligned}\tilde{G}_n(\mathbf{a}) &= |\tilde{g}_n(\mathbf{a})|^2 \\ &= \sum_{k=1}^{N_f}\sum_{\ell=1}^{N_f} \tilde{f}_k(\mathbf{a})\, \tilde{f}_\ell(\mathbf{a})^* \hat{G}(n;k)\hat{G}(n;\ell)^* \\ &+ \sum_{p'=1}^{N_{p'}}\sum_{\ell'=1}^{N_{p'}} i_{p'}(\mathbf{a})\, i_{\ell'}(\mathbf{a})^* \widehat{G'}(n;p')\widehat{G'}(n;\ell')^* \\ &+ 2Re\left\{\sum_{k=1}^{N_f}\sum_{p'=1}^{N_{p'}} \tilde{f}_k(\mathbf{a})\, i_{p'}(\mathbf{a})^* \hat{G}(n;k)\widehat{G'}(n;p')^*\right\} \quad \textbf{Eq. (26)}\end{aligned}$$

This expansion shows that coherent mixing plus square-law intensity readout at the output of the 2nd cascaded diffractive optical processor generates a quadratic feature lift over both previously computed nonlinear fields and newly injected input-encoded complex fields. This creates self-products, input-products, and mixed products that correspond to difference-frequency sets in Fourier space.

If we simplify the diffractive architecture and entirely eliminate the additional input channels at the second SLM plane by choosing $\mathrm{A}_{p'} = 0$, only the first set of cross terms will remain, and Eq. (26) can be simplified as:

$$\tilde{G}_n(\mathbf{a}) = |\tilde{g}_n(\mathbf{a})|^2 = \sum_{k=1}^{N_f}\sum_{\ell=1}^{N_f} \tilde{f}_k(\mathbf{a})\, \tilde{f}_\ell(\mathbf{a})^* \hat{G}(n;k)\hat{G}(n;\ell)^* \quad \textbf{Eq. (27)}$$

In both of these cases described by Eqs. (26) and (27), the intensity detection, $\tilde{G}_n(\mathbf{a}) = |\tilde{g}_n(\mathbf{a})|^2$, at each output channel $n$ of the 2nd (cascaded) diffractive processor includes several cross terms across different complex-valued nonlinear functions implemented by the 1st diffractive processor. These cross terms enhance the overall function representation capabilities of a cascaded diffractive processor, also creating new frequency components not represented at the input illumination plane of the optical processor. Considering a structurally wide and deep diffractive optical processor with large input SLMs for phase encoding, we can have large $N_f$, $N_p$ and $N_{p'}$ values executed in parallel, which can increase the number of available self- and cross-product terms, thereby expanding the expressivity of a field-cascadable diffractive function approximator. This is analogous in spirit to depth-efficiency mechanisms in neural networks, where repeated nonlinear operations can generate increasingly oscillatory or high-order representations. Therefore, this expansion suggests a possible route toward depth-enabled feature growth if nonlinear readout/re-encoding stages are interleaved, but it does not by itself establish Telgarsky-style[32] exponential representation efficiency.

To further expand on this, we analyze the nonlinear function approximation error of this cascaded diffractive architecture following Eq. (26) in relationship to the maximum frequency ($\alpha^{\max} = \max_{\boldsymbol{\alpha}\in\Lambda_c} \| \boldsymbol{\alpha} \|$) content in the target nonlinear function $G_{n,\Lambda_c}(\mathbf{a})$. Here, the cascaded diffractive processor architecture aims to achieve $\tilde{G}_n(\mathbf{a}) \approx G_{n,\Lambda_c}(\mathbf{a})$ for all $n = 1,2,\dots,N_{f,2}$, where $N_{f,2}$ refers to the number of target functions to be simultaneously implemented at the output of the 2nd diffractive processor. Following Eq. (26), the set ($\Lambda_{\text{casc}}$) of frequencies contained in $\tilde{G}_n(\mathbf{a})$ can be written as:

$$\Lambda_{\text{casc}} \subseteq (\Lambda - \Lambda) \cup (\Lambda' - \Lambda') \cup (\Lambda - \Lambda') \cup (\Lambda' - \Lambda) \quad \textbf{Eq. (28)}$$

where $\Lambda = \{\boldsymbol{\alpha}_p\}_{p=1}^{N_p}$ and $\Lambda' = \{\boldsymbol{\alpha}_{p'}\}_{p'=1}^{N_{p'}}$, as defined earlier for the 1st and 2nd input channels, respectively, of the cascaded diffractive processor described by Eq. (26).

If we assume that the maximum encoder frequency at the input of the 1st diffractive processor is given by $\alpha_p^{\max} = \max_{\boldsymbol{\alpha}\in\Lambda} \| \boldsymbol{\alpha} \|$ and similarly, $\alpha_{p'}^{\max} = \max_{\boldsymbol{\alpha}\in\Lambda'} \| \boldsymbol{\alpha} \|$ for the input plane of the 2nd diffractive processor that is cascaded, then the maximum generated computational frequency at the output of the 2nd diffractive processor is bounded by:

$$\alpha_{\text{casc}}^{\max} \leq \max\left\{2\alpha_p^{\max}, 2\alpha_{p'}^{\max}, \alpha_p^{\max} + \alpha_{p'}^{\max}\right\} = 2\max\left\{\alpha_p^{\max}, \alpha_{p'}^{\max}\right\} \quad \textbf{Eq. (29A)}$$

Eqs. (28-29A) represent the spectral support available to the cascaded diffractive architecture. For the simplified cascaded diffractive architecture modeled in Eq. (27) with $A_{p'} = 0$, Eq. (29A) can be written as:

$$\alpha_{\text{casc}}^{\max} \leq 2\alpha_p^{\max} \quad \textbf{Eq. (29B)}$$

Therefore, a necessary spectral condition for accurately approximating a target nonlinear function with a maximum frequency of $\alpha^{\max}$ (i.e., $\tilde{G}_n(\mathbf{a}) \approx G_{n,\Lambda_c}(\mathbf{a})$ based on Eq. (26)) is given by:

$$\alpha^{max} \leq \alpha_{\text{casc}}^{\max} \quad \textbf{Eq. (30)}$$

When this condition of Eq. (30) is satisfied, and when the required spatially varying PSF coefficients can be synthesized with sufficiently small error, we can achieve an accurate function approximation, i.e., $\tilde{G}_n(\mathbf{a}) \approx G_{n,\Lambda_c}(\mathbf{a})$ for all $n = 1,2, \dots, N_{f,2}$ at the output of the cascaded diffractive processor architecture.

To numerically validate this conclusion, we first examined the complex-valued nonlinear functions generated at the outputs of the 1st and 2nd (cascaded) diffractive processors; see Fig. 9. For this analysis, we set $N_f = 100$ complex-valued nonlinear functions at the output of the 1st diffractive processor and $N_{f,2} = 100$ complex-valued nonlinear functions $G_{n,\Lambda_c}(\mathbf{a})$ at the output of the cascaded 2nd diffractive processor, where $n = 1,2, \dots, N_{f,2}$; stated differently, for each value of $\alpha^{\max}$ in Fig. 9, we generated 100 distinct target nonlinear functions for each output plane. The 1st diffractive processor was optimized to approximate its own randomly generated set of $N_f = 100$ complex-valued target nonlinear functions at its output field. It used $N_p = 9$ encoded input channels, with the frequency set $\Lambda = \{\boldsymbol{\alpha}_p\}_{p=1}^{N_p} = \{-4, -3, \dots, 0, \dots, 3, 4\}$ such that $\alpha_p^{\max} = 4$. This processor contained $N \approx 1.25 \times 2 N_p N_f$ trainable phase-only features distributed over $K_L = 4$ diffractive layers. After this first-stage optimization was completed, its diffractive features were fixed and remained unchanged during the subsequent training of the cascaded architecture. At the intermediate plane, $N_{p'} = 8$ additional input channels indexed by $p'$ were introduced with $\Lambda' = \{\boldsymbol{\alpha}_{p'}\}_{p'=1}^{N_{p'}} = \{-8, -7, -6, -5, 5, 6, 7, 8\}$ such that $\alpha_{p'}^{\max} = 8$. Therefore, following Eq. (25), the cascaded 2nd diffractive processor jointly operated on the complex-valued nonlinear fields $\tilde{f}_k(\mathbf{a})$ $(k = 1, \dots, N_f)$ generated by the 1st optimized diffractive processor as well as the additional higher-order Fourier features directly encoded through $\Lambda'$. The 2nd diffractive processor was then independently optimized, while keeping the 1st optimized diffractive processor frozen, to approximate a different set of randomly generated $N_{f,2} = 100$ target nonlinear functions at its output plane. This 2nd diffractive processor contained $N \approx 1.25 \times 2(N_p + N_{p'}) N_{f,2} = 1.25 \times 34 \times N_{f,2}$ trainable phase-only features distributed over $K_L = 4$ diffractive layers.

As shown in Fig. 9a, after its training the 1st diffractive processor's output field maintains a very small error for target nonlinear functions with $\alpha^{\max} \leq \alpha_p^{\max} = 4$, whereas the approximation error increases significantly once $\alpha^{\max} > 4$. This transition is also evident from representative cases ① and ② in Fig. 9b. In comparison, the cascaded diffractive processor's output remains highly accurate up to $\alpha^{\max} \approx \alpha_{p'}^{\max} = 8$ and begins to fail when the target functions' $\alpha^{\max}$ exceeds 8; see cases ③ and ④ in Fig. 9b. These results confirm that, for complex optical field-based function approximation, the additional intermediate input encoder channels can be used to extend the available frequency support of output functions from that provided by the first input encoding plane. This extension is enabled by the injection of additional encoding features at the intermediate plane, following Eq. (25).

We next analyzed the optical intensity-based nonlinear function approximation following Eq. (26). As shown in Fig. 10, for this numerical analysis, we also used $N_{f,2} = 100$ distinct target nonlinear functions $G_{n,\Lambda_c}(\mathbf{a})$ for each value of $\alpha^{\max}$, while assuming $N_f = 100$ for the 1st diffractive processor, same as in Fig. 9. Following the analysis in Fig. 9, the first input frequency set was selected as $\Lambda = \{-4, -3, \dots 0, \dots, 3, 4\}$, with $N_p = 9$ and $\alpha_p^{\max} = 4$, whereas the additional intermediate frequency set was $\Lambda' = \{-8, -7, -6, -5, 5, 6, 7, 8\}$, with $N_{p'} = 8$ and $\alpha_{p'}^{\max} = 8$. The 2nd diffractive processor here has $N = 1.25 \times 2(N_p + N_{p'})N_{f,2} = 1.25 \times 34 \times N_{f,2}$ optimizable phase features.

Based on Eq. (29A), these choices set: $\alpha_{\text{casc}}^{\max} \le 2\alpha_{p'}^{\max} = 16$. Our results, shown in Fig. 10, confirm Eqs. (29A, 30), i.e., $\alpha^{max} \le \alpha_{\text{casc}}^{\max} \le 16$ is a necessary condition for accurately approximating target nonlinear functions ($\tilde{G}_n(\mathbf{a}) \approx G_{n,\Lambda_c}(\mathbf{a})$ for all $n = 1, 2, \dots, N_{f,2}$ ) each with a maximum frequency of $\alpha^{\max}$. Figure 10a supports that at the output plane of the 2nd diffractive processor, we maintain a relatively small MSE for target nonlinear functions when $\alpha^{\max} \le 2\alpha_{p'}^{\max} = 16$. As shown in Fig. 10b, the examples in cases ② and ④ further show that accurate approximation is retained near the predicted spectral boundary, while noticeable deviations/errors emerge once $\alpha^{\max} > 2\alpha_{p'}^{\max} = 16$, further validating Eqs. (29A, 30).

For comparison, Fig. 10 also shows the intensity-based function approximation output of a single diffractive processor using $N_p = 17$ input encoder channels with $\Lambda = \{-8, -7, \dots 0, \dots, 7, 8\}$, such that $\alpha_p^{\max} = 8$ . This single-processor design used $N \approx 1.25 \times 2N_pN_{f,2} = 1.25 \times 34 \times N_{f,2}$ optimized phase-only features distributed over $K_L = 4$ diffractive layers. In agreement with the simplified spectral bound reported in Eq. (29B), this single diffractive processor configuration also remains accurate in its function approximations when $\alpha^{\max} \le 2\alpha_p^{\max} = 16$, after which the approximation error increases, as illustrated by representative cases shown in ① and ③. Importantly, although both diffractive architectures shown in Fig. 10a have the same trainable degrees of freedom ($N \approx 1.25 \times 34 \times N_{f,2}$) and exhibit a similar spectral cutoff near $\alpha^{\max} = 16$, the cascaded diffractive processor achieves a lower function approximation error across the evaluated high frequency range, especially for $\alpha^{\max} > 5$. This improvement highlights that the intermediate encoding and the 2nd diffractive transformation through the cascaded optical processor provide a richer set of optical features for synthesizing the target intensity functions, as indicated in Eqs. (26), (28), and (29A).

Taken together, Figs. 9 and 10 numerically support our analyses developed in Eqs. (25)–(30). At the complex optical field level, the accessible frequencies remain limited by those directly provided through the input encodings. Importantly, the intermediate encoding through $\Lambda'$ can be used to supply new higher-order frequency components that are absent from the first-stage input set $\Lambda$, thereby expanding the feature space available to the cascaded diffractive processor. Following this intermediate feature injection, intensity detection generates additional difference-frequency components through the quadratic self- and cross-product terms reported in Eqs. (26, 28). These results support a possible route toward depth-enabled feature growth using intermediate feature injection and optical intensity readout.

Overall, architectural depth is significant to enhance the performance of diffractive optical processors. Additional diffractive layers can improve physical wave mixing, reduce ballistic photon components at the output, increase controllability and accuracy of the estimated PSFs (which dictate function approximation error bound, $E_{\mathrm{PSF}}$, as detailed in earlier sections), improve rank utilization within the available spatial modes, and improve the accuracy-efficiency trade-off observed in earlier works[11,17,29,30,33–35]. Earlier studies also demonstrated both theoretically and empirically various depth advantages of diffractive processors and quantified the necessary conditions to achieve $E_{\mathrm{PSF}} \approx 0$.[15–18]

These are important gains in optical expressivity, but they are not mathematically identical to the abstract depth-separation theorems for digital neural networks and do not prove Telgarsky-style depth separation for a diffractive optical processor architecture. Stated differently, establishing mathematically complete optical analogs of depth-separation results remains an open problem for diffractive optical processors.

Finally, we note that if the final cascaded output is used to represent signed or complex-valued functions, the intensity outputs, $|\tilde{g}_n(\mathbf{a})|^2$, must again be decoded using, e.g., differential or multi-channel readout conventions that we introduced in earlier sections; see e.g., the paragraph of Eq. (13).

**Coherence regimes and spatially incoherent intensity-domain Fourier-feature approximation**

The framework analyzed above focuses primarily on coherent phase-encoded diffractive function approximation, where the computational Fourier features are encoded as complex fields, $u_{\mathrm{in}}(p;\mathbf{a}) = \exp\left(j2\pi\boldsymbol{\alpha}_p \cdot \mathbf{a}\right)$, and the diffractive processor synthesizes complex-valued coefficients $\hat{F}(k;p)$ through spatially varying coherent PSFs. Recent results on nonlinear optical computing with incoherent and partially coherent light show that a complementary intensity-domain version of the same Fourier-feature principle is also possible[12]. In such an architecture that is spatially incoherent or partially coherent, the input variable is encoded into nonnegative intensity features[12]. In the case of spatially incoherent illumination, the diffractive processor optically synthesizes the desired output nonlinear functions through spatially varying incoherent[26,36] intensity PSFs followed by differential intensity readout[12,37]. Following the notation of our analyses in earlier sections, a spatially incoherent diffractive function approximator uses cosine and sine intensity encodings at its input plane, written as:

$$I_{\cos}(p;\mathbf{a}) = \frac{1+\cos\left(2\pi\boldsymbol{\alpha}_p \cdot \mathbf{a}\right)}{2}, \quad I_{\sin}(p';\mathbf{a}) = \frac{1+\sin\left(2\pi\boldsymbol{\alpha}_{p'} \cdot \mathbf{a}\right)}{2} \qquad \mathbf{Eq.\ (31)}$$

where, in this sub-section, $p$ and $p'$ index the input encoder plane with two spatially separate regions, one dedicated for $I_{\cos}(p;\mathbf{a})$ and one dedicated for $I_{\sin}(p';\mathbf{a})$. For these 2 regions, we also have $\boldsymbol{\alpha}_p \in \Lambda_i = \{\boldsymbol{\alpha}_p\}_{p=1}^{N_p}$ and $\boldsymbol{\alpha}_{p'} \in \Lambda'_i = \{\boldsymbol{\alpha}_{p'}\}_{p'=1}^{N_{p'}}$, each of which defines a nonzero half-spectrum. For simplicity, we assume $N_p = N_{p'}$ and $\Lambda_i = \Lambda'_i$. These $2N_p$ nonnegative intensity channels are the incoherent analog of the $N_p$ coherent complex

Fourier features used in Eqs. (2-3). With this notation, a spatially incoherent (inc) intensity-based diffractive approximator for the output channel $k$ (with a differential detection scheme[12]) can be written as:

$$\hat{f}_{k_+}^{\mathrm{inc}}(\mathbf{a}) = \sum_{p=1}^{N_p} \widehat{W}_{\cos}(k_+;p)\, I_{\cos}(p;\mathbf{a}) + \sum_{p\prime=1}^{N_p} \widehat{W}_{\sin}(k_+;p')\, I_{\sin}(p';\mathbf{a})\,;\, k_+ = 1,\dots,N_f$$

$$\hat{f}_{k_-}^{\mathrm{inc}}(\mathbf{a}) = \sum_{p=1}^{N_p} \widehat{W}_{\cos}(k_-;p)\, I_{\cos}(p;\mathbf{a}) + \sum_{p\prime=1}^{N_p} \widehat{W}_{\sin}(k_-;p')\, I_{\sin}(p';\mathbf{a})\,;\, k_- = 1,\dots,N_f \quad \mathbf{Eq.\,(32)}$$

For the differential signal, we have:

$$\begin{aligned}\hat{f}_{k}^{\mathrm{inc}}(\mathbf{a}) &= \hat{f}_{k_+}^{\mathrm{inc}}(\mathbf{a}) - \hat{f}_{k_-}^{\mathrm{inc}}(\mathbf{a}) \\ &= \frac{1}{2}\sum_{p=1}^{N_p} \widehat{W}_{\cos}(k;p)\left(1+\cos\left(2\pi\boldsymbol{\alpha}_p\cdot\mathbf{a}\right)\right) \\ &+ \frac{1}{2}\sum_{p\prime=1}^{N_p} \widehat{W}_{\sin}(k;p')\left(1+\sin\left(2\pi\boldsymbol{\alpha}_{p\prime}\cdot\mathbf{a}\right)\right)\end{aligned}$$

Here $\widehat{W}_{\cos}(k;p) = \widehat{W}_{\cos}(k_+;p) - \widehat{W}_{\cos}(k_-;p)$ and $\widehat{W}_{\sin}(k;p') = \widehat{W}_{\sin}(k_+;p') - \widehat{W}_{\sin}(k_-;p')$. Furthermore, $\widehat{W}_{\cos}(k_+;p) \geq 0$, $\widehat{W}_{\cos}(k_-;p) \geq 0$, $\widehat{W}_{\sin}(k_+;p') \geq 0$ and $\widehat{W}_{\sin}(k_-;p') \geq 0$ refer to the set of spatially varying incoherent PSFs that are jointly implemented by the same diffractive optical processor[26,36]. To keep these 4 sets of spatially varying incoherent PSFs distinct from each other, we keep both $p$ and $p'$ indices even if we assumed earlier $N_p = N_{p\prime}$ and $\Lambda_i = \Lambda_i'$.

The above equation can be further simplified as:

$$\hat{f}_{k}^{\mathrm{inc}}(\mathbf{a}) = \hat{C}_k + \sum_{p=1}^{N_p} \hat{A}_k\left(\boldsymbol{\alpha}_p\right)\cos\left(2\pi\boldsymbol{\alpha}_p\cdot\mathbf{a}\right) + \sum_{p\prime=1}^{N_p} \hat{B}_k\left(\boldsymbol{\alpha}_{p'}\right)\sin\left(2\pi\boldsymbol{\alpha}_{p\prime}\cdot\mathbf{a}\right), \quad \mathbf{Eq.\,(33)}$$

where we have:

$$\hat{A}_k\left(\boldsymbol{\alpha}_p\right) = \tfrac{1}{2}\widehat{W}_{\cos}(k;p),\ \hat{B}_k\left(\boldsymbol{\alpha}_{p'}\right) = \tfrac{1}{2}\widehat{W}_{\sin}(k;p'),\ \hat{C}_k = \tfrac{1}{2}\sum_{p=1}^{N_p}\widehat{W}_{\cos}(k;p) + \tfrac{1}{2}\sum_{p\prime=1}^{N_p}\widehat{W}_{\sin}(k;p')$$

Therefore, under spatially incoherent illumination, the optimized diffractive processor performs finite Fourier-feature coefficient synthesis, but the coefficient synthesis occurs in the intensity domain rather than in the coherent complex-field domain. The target (i.e., ground truth) real-valued Fourier approximation can be written as:

$$f_{k,\Lambda_i}^{\mathrm{real}}(\mathbf{a}) = C_k + \sum_{p=1}^{N_p} \left[ A_k(\boldsymbol{\alpha}_p)\cos(2\pi\boldsymbol{\alpha}_p \cdot \mathbf{a}) + B_k(\boldsymbol{\alpha}_p)\sin(2\pi\boldsymbol{\alpha}_p \cdot \mathbf{a}) \right] \ \textbf{Eq. (34)}$$

The corresponding spatially incoherent intensity-PSF coefficient error satisfies the conservative uniform bound, given by **Eq. (35)**:

$$\| f_{k,\Lambda_i}^{\mathrm{real}} - \hat{f}_k^{\mathrm{inc}} \|_{\infty,K} \leq | C_k - \hat{C}_k | + \sum_{p=1}^{N_p} \left[ \left| A_k(\boldsymbol{\alpha}_p) - \hat{A}_k(\boldsymbol{\alpha}_p) \right| + \left| B_k(\boldsymbol{\alpha}_p) - \hat{B}_k(\boldsymbol{\alpha}_p) \right| \right] = E_{\mathrm{IPSF},k}$$

which is an incoherent analog of the coherent PSF-synthesis error $E_{\mathrm{PSF},k}$, defined in Eq. (5). In practice, a baseline subtraction step can be used to suppress the effective DC mismatch, i.e., the term $| C_k - \hat{C}_k |$, and the $E_{\mathrm{IPSF},k}$ error bound in Eq. (35) can be reduced to the sine/cosine coefficient-synthesis terms shown above.

This spatially incoherent construction reinforces the central interpretation of diffractive nonlinear computation as the synthesis of physical Fourier-feature coefficients. However, it also highlights an important distinction: because optical intensity is real and nonnegative, arbitrary signed Fourier coefficients cannot be represented by a single nonnegative intensity channel alone. They require an explicit decoding convention, such as differential detection, baseline offsets, or multiple detector channels. Following the notation of Eq. (13), as an example, one can have $g_k^{\mathrm{inc}}(\mathbf{a}) = \sum_{x=0}^{2} w_x \hat{f}_{k,x}^{\mathrm{inc}}(\mathbf{a})$ with $w_x = \exp\left(jx\frac{2\pi}{3}\right)$ to approximate any complex-valued nonlinear function (after appropriate normalization and nonnegative channel decomposition) using 3 intensity readout values ($\hat{f}_{k,x=0,1,2}^{\mathrm{inc}}$) at the output of a spatially incoherent diffractive processor. The incoherent processor implements this by assigning each target function to a detector region and decoding the scalar output through a weighted combination ($w_x$) of the measured intensities.

The incoherent diffractive processor architecture also gives a useful resource-scaling comparison. In the scalar implementation, $2N_p$ input intensity channels are used for cosine/sine feature encoding, and a differential[12] readout scheme with 2 detectors (i.e., $k_+$ and $k_-$ in Eq. (32)) for each function $k$ gives an effective output dimension proportional to $2N_f$. Therefore, a spatially incoherent[26,36] diffractive processor with phase-only degrees of freedom used for nonlinear function approximation has a scaling of $N \gtrsim 2N_iN_o = 8N_pN_f$. This is consistent with the broader conclusion of the present work: massive output parallelism does not remove the need for optical degrees of freedom that scale with the product of the input spectral feature dimension and the number of output functions.

Finally, Eqs. (31)–(35) describe the spatially incoherent illumination case. Under partially coherent illumination, the same architecture can be analyzed using a coherence-aware statistical forward model[12,38].

## Discussion

Diffractive optical processors provide a physically distinctive realization of universal function approximation through passive linear wave propagation combined with nonlinear phase encoding of the input variables. Rather than implementing nonlinear activations within successive computational layers, these architectures realize finite Fourier-feature expansions whose coefficients are synthesized by optimized spatially varying coherent PSFs. This establishes a direct connection between classical universal approximation theory, harmonic analysis, and diffractive optical computing, placing recent demonstrations of massively parallel nonlinear function approximation within a rigorous mathematical framework.

This interpretation also extends beyond the coherent phase-encoded setting: passive diffractive processors can approximate nonlinear functions under spatially incoherent or partially coherent illumination when the input variables are encoded into nonnegative intensity features such as $[1 + \cos(2\pi\boldsymbol{\alpha}_p \cdot \mathbf{a})]/2$ and $[1 + \sin(2\pi\boldsymbol{\alpha}_p \cdot \mathbf{a})]/2$, followed by spatially varying incoherent intensity PSFs and differential intensity readout[12]. In this incoherent regime, the processor again realizes a finite Fourier-feature expansion, but the learned coefficients are synthesized in the intensity domain rather than the coherent complex-field domain. This complementary result strengthens the central conclusion that the nonlinear dependence on the computational variable $\mathbf{a}$ is introduced by the input encoding and readout model, while the passive diffractive processor volume supplies a learned linear optical coefficient map created by spatially varying PSFs.

It is important to distinguish the present framework from the classical use of Fourier optics in computational imaging and optical information processing. Although both employ coherent optical propagation and Fourier representations, they operate on fundamentally different mathematical objects. Classical Fourier optics manipulates the spatial-frequency content of propagating electromagnetic wavefields. Its Fourier variables $(f_x, f_y, f_z)$ represent the angular spectrum of the optical field and therefore correspond directly to physical wave vectors, propagation angles, diffraction, and imaging. Operations such as Fourier transformation, spatial filtering, convolution, correlation, holography, and image formation all manipulate these physically propagating spatial-frequency components.

By contrast, the computational Fourier variables $\boldsymbol{\alpha}_p$ employed in the present framework are basis indices associated with an abstract mathematical variable $\mathbf{a} \in \mathbb{R}^D$, introduced solely to construct a Fourier representation of the $N_f$ target nonlinear functions implemented optically. They do not represent optical wave vectors, propagation angles, diffraction orders, or any physically propagating spatial-frequency components of the electromagnetic field. Furthermore, unlike conventional Fourier optics, the dimensionality $D$ of the computational variable is determined by the function-approximation problem rather than by physical space and is therefore not restricted to three dimensions. An arbitrarily large $D \gg 1$ can be considered, although an accurate approximation generally requires correspondingly larger Fourier-feature sets (larger $N_p$). These computational Fourier frequencies are prescribed during the input phase-encoding stage through $u_{\text{in}}(p; \mathbf{a}) = \exp\left(j2\pi\, \boldsymbol{\alpha}_p \cdot \mathbf{a}\right)$ before optical propagation begins. The subsequent diffractive processor does not compute a Fourier transform of the incident optical field; instead, it synthesizes the coefficients of a prescribed finite Fourier expansion through its optimized spatially varying coherent PSFs. In the present framework, optical propagation serves only as the physical mechanism for implementing learned linear combinations of pre-encoded Fourier features; the Fourier frequencies themselves are conceptual computational variables rather than physical spatial frequencies. Consequently, the proposed architecture should not be regarded as an extension of conventional Fourier optics or spatial-frequency filtering, but rather as a programmable optical realization of finite Fourier-feature approximation for nonlinear function computation, consistent with the wavefront-encoding framework of this work.

It is also important to emphasize that the mathematical existence of a universal approximator should not be conflated with its practical physical realization. Classical universal approximation theorems establish only the existence of finite representations in suitable function spaces, whereas practical optical implementations via diffractive optical processors are further constrained by finite Fourier bandwidth, finite optical degrees of freedom, diffraction efficiency, fabrication tolerances, detector noise, and hardware quantization. Consequently, the total approximation error of a diffractive function approximator is governed not only by the Fourier truncation error ($E_{\text{trunc}}$), coefficient-synthesis error ($E_{\text{PSF}}$) and input phase-related error ($E_{\phi}$), but also by implementation-dependent hardware and measurement errors, summarized in $E_{\text{misc}}$. These physical constraints determine the achievable approximation accuracy for any practical diffractive processor and therefore must be analyzed independently of classical UFA existence theorems.

The presented framework also clarifies the relationship between approximation theory and statistical learning from the perspective of diffractive optical processors. Universal approximation establishes representational capability, whereas learnability concerns the number of training samples required to identify a suitable optical realization from finite data. The finite-class sample-complexity analysis presented here demonstrates that, under quantized phase modulation, the required number (M) of training samples depends jointly on the number of trainable optical degrees of freedom ($N$), phase-bit depth ($B$), tolerated generalization gap ($E$), and desired confidence level dictated by $1-\delta$. Accordingly, statistical generalization properties cannot be inferred solely from representational universality, but instead require explicit capacity-control analyses for the physically realizable diffractive hypothesis class.

From a systems perspective, this work also highlights the interplay between approximation complexity and physical optical resources. Although diffractive processors can simultaneously compute extremely large numbers of nonlinear functions with diffraction-limited spatial density and ultrafast latency, they do not remove the worst-case curse of dimensionality. The required number of Fourier features, input encoding channels, and trainable optical degrees of freedom continues to scale with the complexity of the target function class. Consequently, optical space-bandwidth product, diffraction efficiency, wavelength[17,39]/polarization[16,40]/phase multiplexing[19], detector count, and available trainable phase features become fundamental computational resources that jointly determine achievable approximation performance.

Our coherent-cascade analysis further reveals that field-level optical cascadability provides a mechanism for generating increasingly rich feature representations through coherent interference and subsequent optical intensity readout. The resulting feature expansion, detailed in our Results section, provides a potential route toward progressively richer optical representations across multiple stages. Nevertheless, these mechanisms should not be interpreted as establishing the depth-separation theorems known for digital neural networks. Although coherent optical cascades can substantially enhance representational richness, establishing mathematically complete optical analogs of depth-separation results remains an open problem.

More broadly, the framework developed here bridges mathematical approximation theory, statistical learning theory, and physical optical implementation. It provides a common language for analyzing how approximation accuracy, learnability, optical throughput, hardware complexity, and photon efficiency jointly determine the performance limits of diffractive function approximators. These results suggest that future developments in analog optical computing should be guided not only by representational expressivity but also by quantitative analyses of physical resource allocation, statistical generalization, and hardware-aware optimization. Such a unified theoretical perspective may facilitate the development of increasingly scalable diffractive computing architectures capable of performing large-scale nonlinear computation with high parallelism, low latency, and improved energy efficiency.

**Supplementary Information** includes:

- **Optical forward model under spatially coherent illumination**
- **Optical forward model under spatially incoherent illumination**

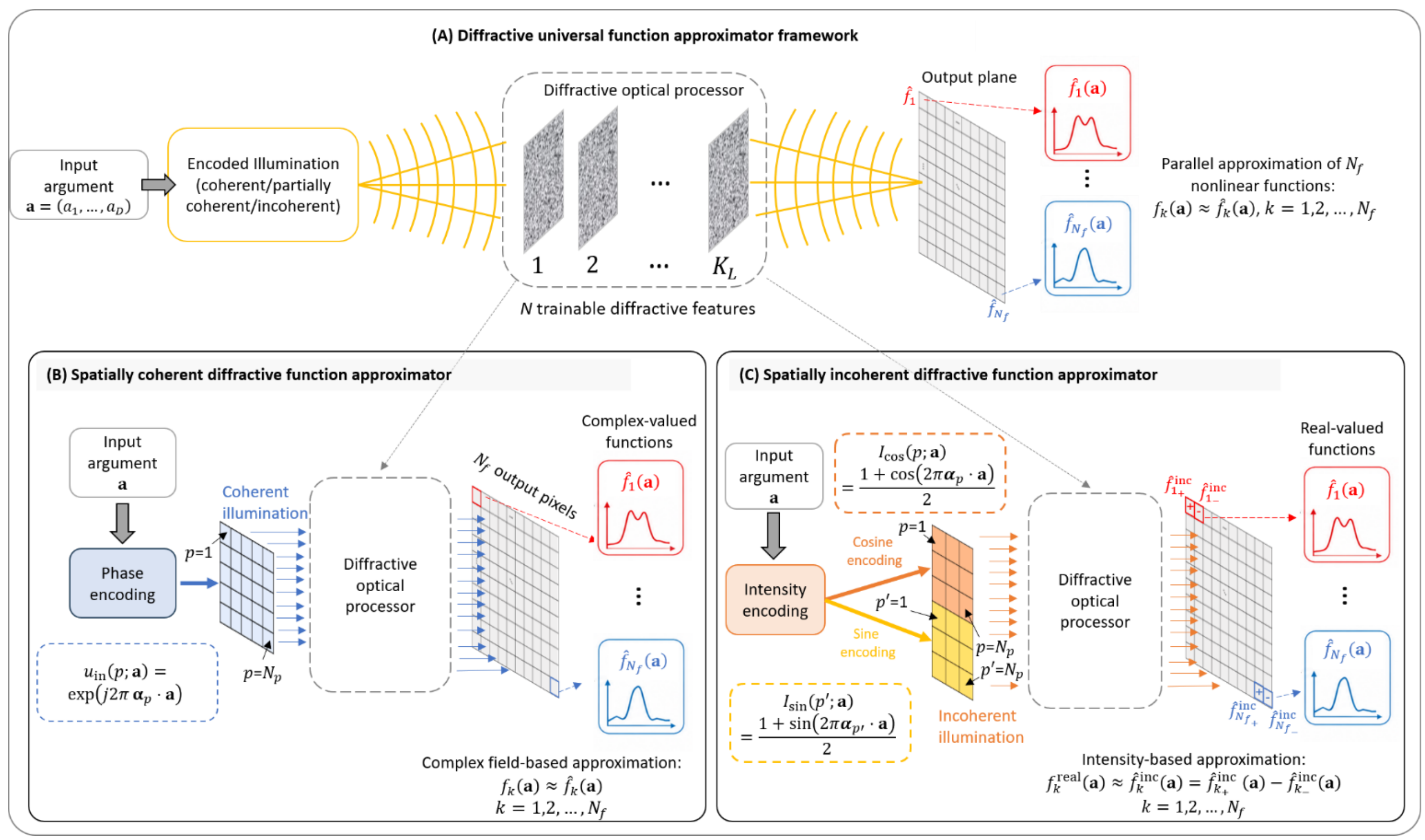

**Fig. 1 | Universal function approximation using diffractive optical processors. (a)** Conceptual overview of a diffractive universal function approximator. An input argument $\mathbf{a} = (a_1, \dots, a_D)$ is encoded into the incident optical illumination and propagates through a $K_L$-layer diffractive optical processor comprising $N$ trainable diffractive features. The resulting optical field or intensity distribution at the output plane provides parallel estimates $\hat{f}_1(\mathbf{a}), \dots, \hat{f}_{N_f}(\mathbf{a})$ of $N_f$ target nonlinear functions. **(b)** Schematic of a diffractive function approximator under spatially coherent illumination. The input argument is phase encoded across $N_p$ input pixels according to $u_{\text{in}}(p; \mathbf{a}) = \exp\left(j2\pi\boldsymbol{\alpha}_p \cdot \mathbf{a}\right)$. Under coherent illumination, the optimized diffractive processor maps the encoded wavefront to $N_f$ designated output pixels, where the complex field at each output pixel approximates the target nonlinear function, i.e., $f_k(\mathbf{a}) \approx \hat{f}_k(\mathbf{a})$, executed all in parallel for $k = 1, \dots, N_f$. **(c)** Schematic of a diffractive function approximator under spatially incoherent illumination, which can also be used for partially coherent illumination. The input argument is represented by nonnegative cosine and sine intensity encodings. Output intensity values are grouped into positive and negative channels, and their differential combinations yield $N_f$ real-valued nonlinear function estimation, i.e., $\hat{f}_k^{\text{inc}}(\mathbf{a}) = \hat{f}_{k+}^{\text{inc}}(\mathbf{a}) - \hat{f}_{k-}^{\text{inc}}(\mathbf{a})$, executed all in parallel for $k = 1, \dots, N_f$.

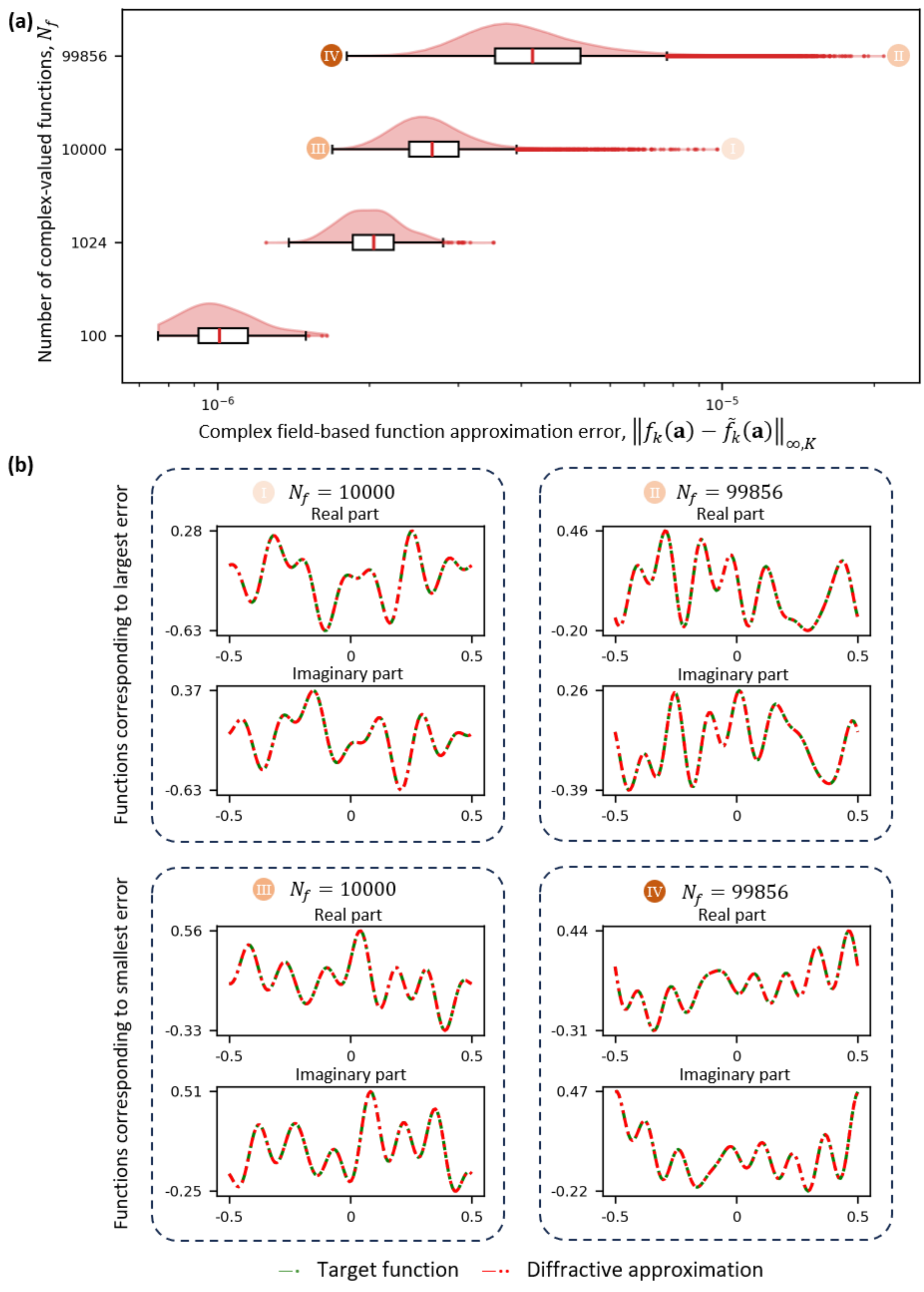

**Fig. 2 | Accuracy of coherent diffractive optical processors for increasing numbers of**

**target complex-valued nonlinear functions.** (a) Distribution of function approximation errors for four diffractive processor designs corresponding to different values of $N_f$ (number of target nonlinear functions), ranging from 100 to ~100,000. Each design uses $N \approx 1.25 \times 2N_pN_f$ optimized and fixed phase-only features distributed over $K_L = 2$ diffractive surfaces, where $N_p = 9$ is the number of input pixels used to encode the argument $a = [-0.5 : 0.5]$ of the functions. As $N_f$ increases, the spread and upper bound of the error distribution also increase relatively; however, the error remains negligible across all nonlinear functions, supporting the scalability of the architecture. (b) Examples of the nonlinear functions with the maximum and minimum errors from the two largest designs ($N_f = 10^4$ and $N_f \approx 10^5$). Real and imaginary parts of the target functions (green curves) and their diffractive approximations (red curves) are plotted over $a = [-0.5 : 0.5]$, showing very good agreement between the two.

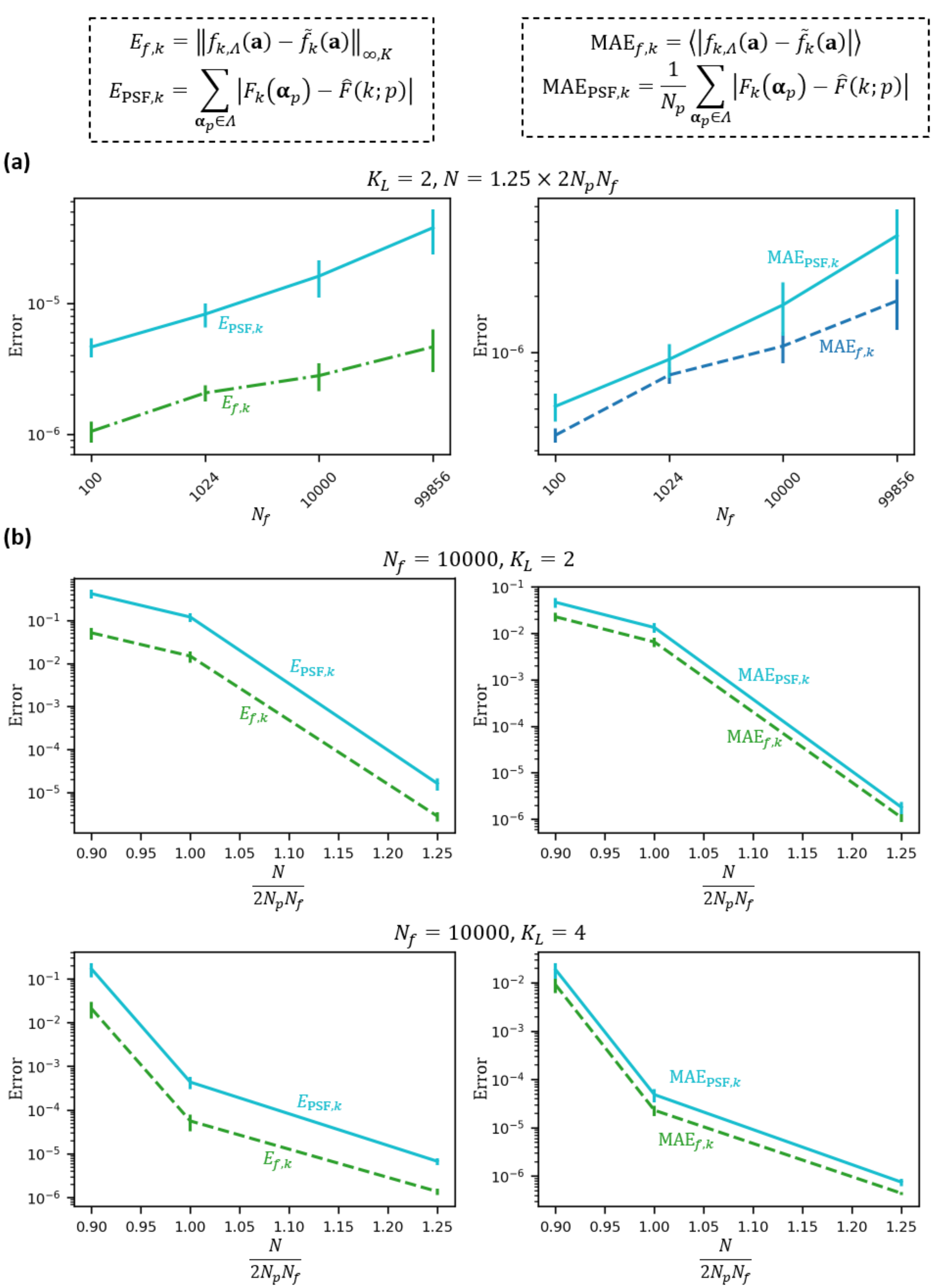

**Fig. 3 | Numerical analysis of complex field-based function-approximation and PSF-synthesis errors.** The curves and error bars represent the mean and standard deviation,

respectively, of each error metric $E_{f,k}$, $E_{\mathrm{PSF},k}$, $\mathrm{MAE}_{f,k}$ and $\mathrm{MAE}_{\mathrm{PSF},k}$ over $N_f$ target nonlinear functions. **(a)** $E_{f,k}$, and $E_{\mathrm{PSF},k}$ (left), and $\mathrm{MAE}_{f,k}$ and $\mathrm{MAE}_{\mathrm{PSF},k}$(right), as a function of $N_f$. Although the errors increase with $N_f$, they remain negligible even for $N_f \sim 10^5$ as the number of trainable diffractive features $N$ exceeds $2N_pN_f$. Here, $N = 1.25(2N_pN_f)$. **(b)** The same error metrics as a function of $N/(2N_pN_f)$ for $N_f = 10{,}000$ nonlinear functions implemented in parallel. Results are shown for $K_L = 2$ (top) and $K_L = 4$ (bottom) diffractive layers. The errors can be substantial when $N < 2N_pN_f$, but become negligible once $N$ exceeds $2N_pN_f$. Also, for the same $N$, distributing the trainable diffractive features over a larger number of diffractive layers further reduces the errors.

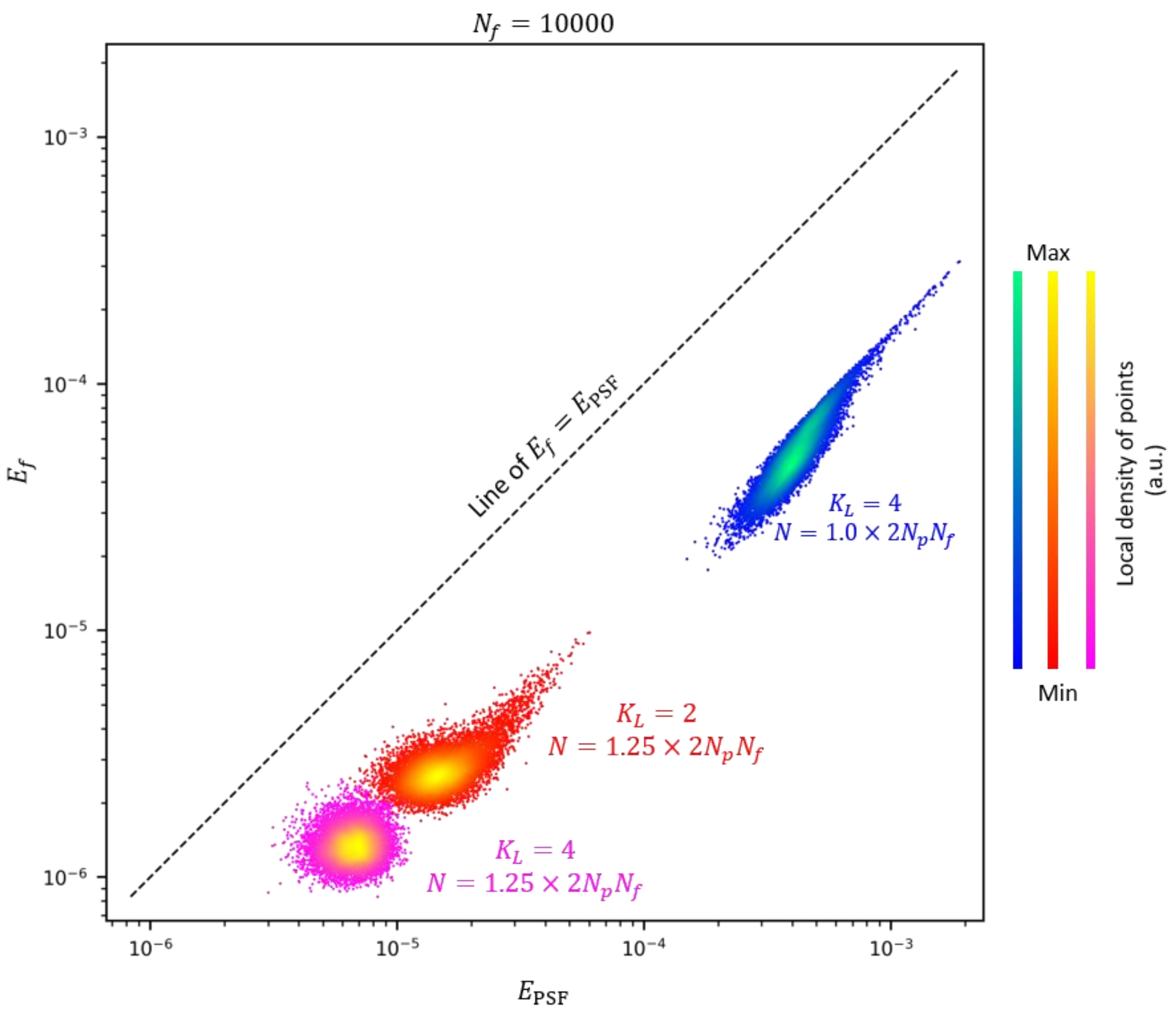

**Fig. 4 | Function-approximation maximum error vs. PSF approximation error for individual output nonlinear functions.** Scatter plots of $E_{f,k}$versus $E_{\mathrm{PSF},k}$ for $k = 1, \ldots, N_f$, with $N_f = 10{,}000$, are shown for three distinct diffractive processor designs, where the number of diffractive layers, $K_L$, and the trainable diffractive features, $N$, vary. Each point represents one approximated function and is colored according to the local point density. All sampled points lie below the line $E_f = E_{\mathrm{PSF}}$, consistent with the conservative bound in Eq. (5). Increasing $N$ from $2N_pN_f$ to $1.25(2N_pN_f)$ substantially reduces errors, while

distributing the same number of diffractive features over more layers ($K_L = 4$ vs. $K_L = 2$) provides a further reduction in error.

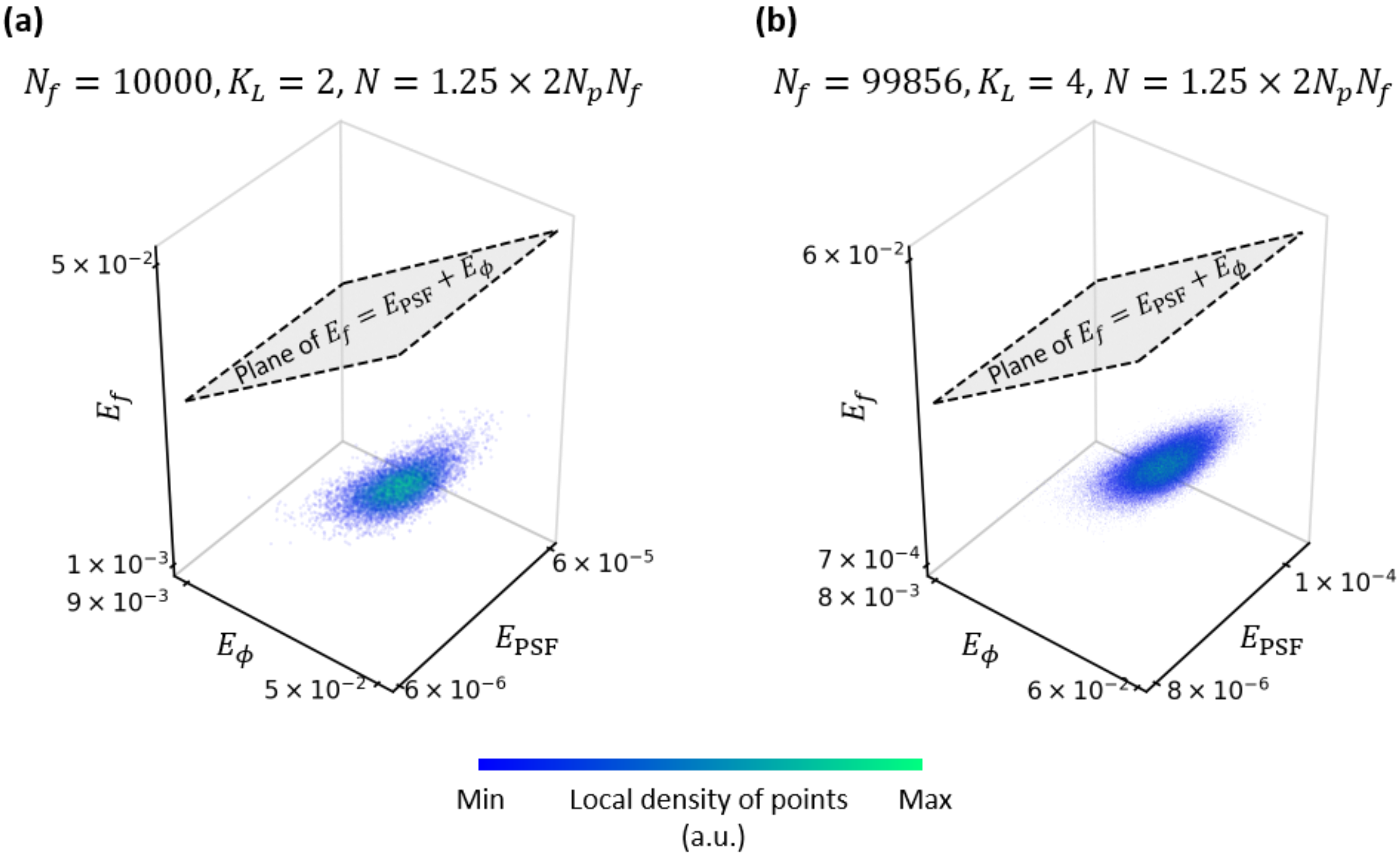

**Fig. 5 | Function-approximation maximum error vs. PSF error and input phase error for individual output nonlinear functions.** Three-dimensional scatter plots of $E_{f,k}$, $E_{\text{PSF},k}$, and $E_{\phi,k}$ for the individual approximated nonlinear functions are shown for $N_f = 10{,}000$, $K_L = 2$, and $N = 1.25 \times 2N_pN_f$ **(a)** and $N_f = 99{,}856$, $K_L = 4$, and $N = 1.25 \times 2N_pN_f$ (**b**). Each point represents one output function and is colored according to the local point density. In both cases, all points remain below the plane $E_f = E_{\text{PSF}} + E_\phi$, being consistent with the conservative error bound in Eq. (7).

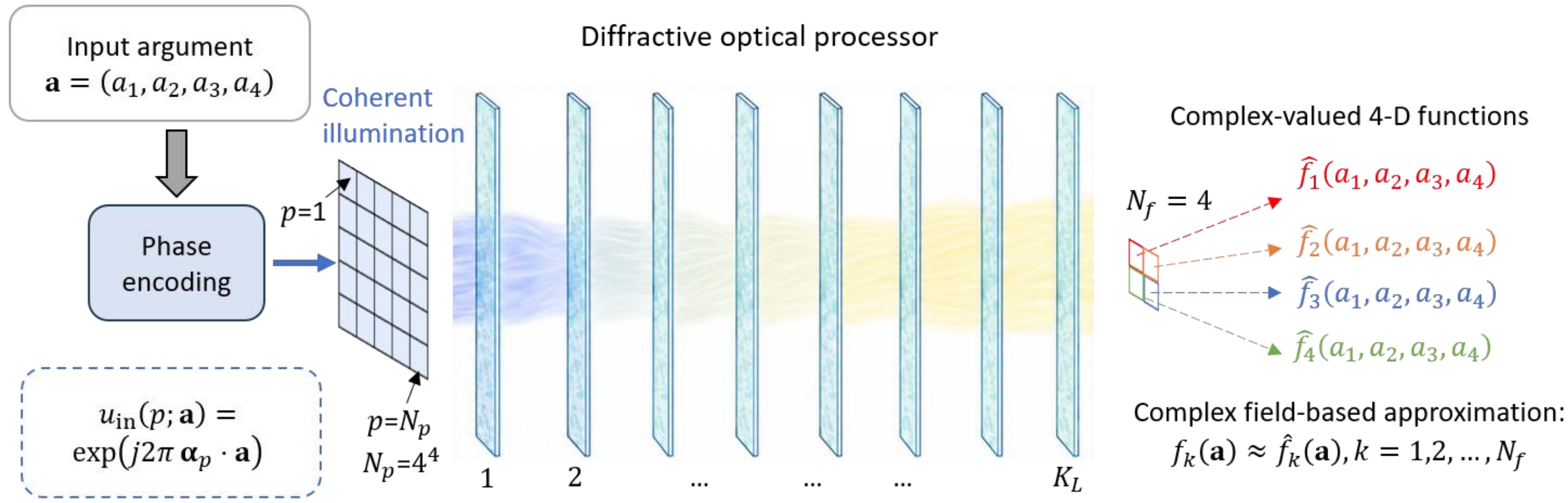

**Fig. 6 | Complex-valued 4-D nonlinear function approximation using a diffractive optical processor.** Schematic of the diffractive optical processor used to approximate four independent complex-valued 4-D nonlinear functions, i.e., $f_1(a_1, a_2, a_3, a_4)$, $f_2(a_1, a_2, a_3, a_4)$, $f_3(a_1, a_2, a_3, a_4)$, and $f_4(a_1, a_2, a_3, a_4)$. A $16 \times 16$ phase-encoded input aperture, corresponding to $N_p = F^D = 4^4 = 256$, is followed by $K_L = 8$ phase-only diffractive layers that are spatially optimized for this 4-D nonlinear function approximation task. The output plane contains $N_f = 2 \times 2$ output pixels, one for each nonlinear function; the width of each output pixel is $\lambda/2$. The complex field at these output pixels reveals: $\hat{f}_k(a_1, a_2, a_3, a_4) \approx f_k(a_1, a_2, a_3, a_4)$ for $k = 1, .., N_f$, with $N_f = 4$ shown in this case (also see Figs. 7-8).

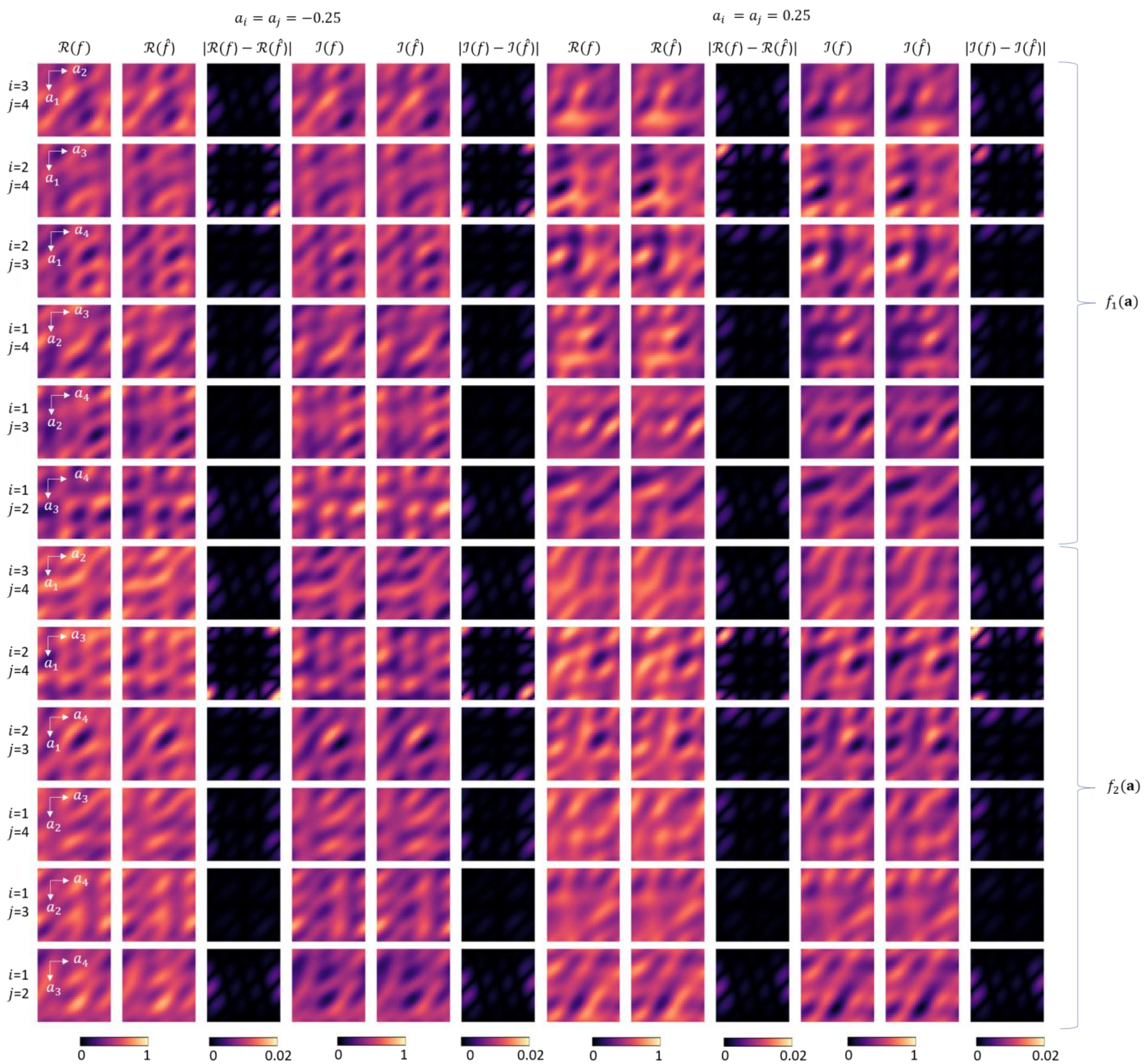

**Fig. 7 | 2-D slices of complex-valued 4-D nonlinear functions, comparing the ground truth nonlinear functions $f_1(a_1, a_2, a_3, a_4)$ and $f_2(a_1, a_2, a_3, a_4)$ with their all-optical approximations $\hat{f}_1(a_1, a_2, a_3, a_4)$ and $\hat{f}_2(a_1, a_2, a_3, a_4)$, respectively.** Each row shows one unique slice, where two coordinates are fixed to a common value, and the remaining two coordinates form the displayed 2-D plane. Within each group, the panels show the target nonlinear function, diffractive processor output, and absolute error for both the real ($\mathcal{R}$) and imaginary ($\mathcal{I}$) parts of each function. $N_p = F^D = 4^4 = 256$, and $K_L = 8$; see Fig. 6. The 4-D MAE values, computed over the full 4-D validation grid, are 0.002814 for $f_1$ and 0.003309 for $f_2$.

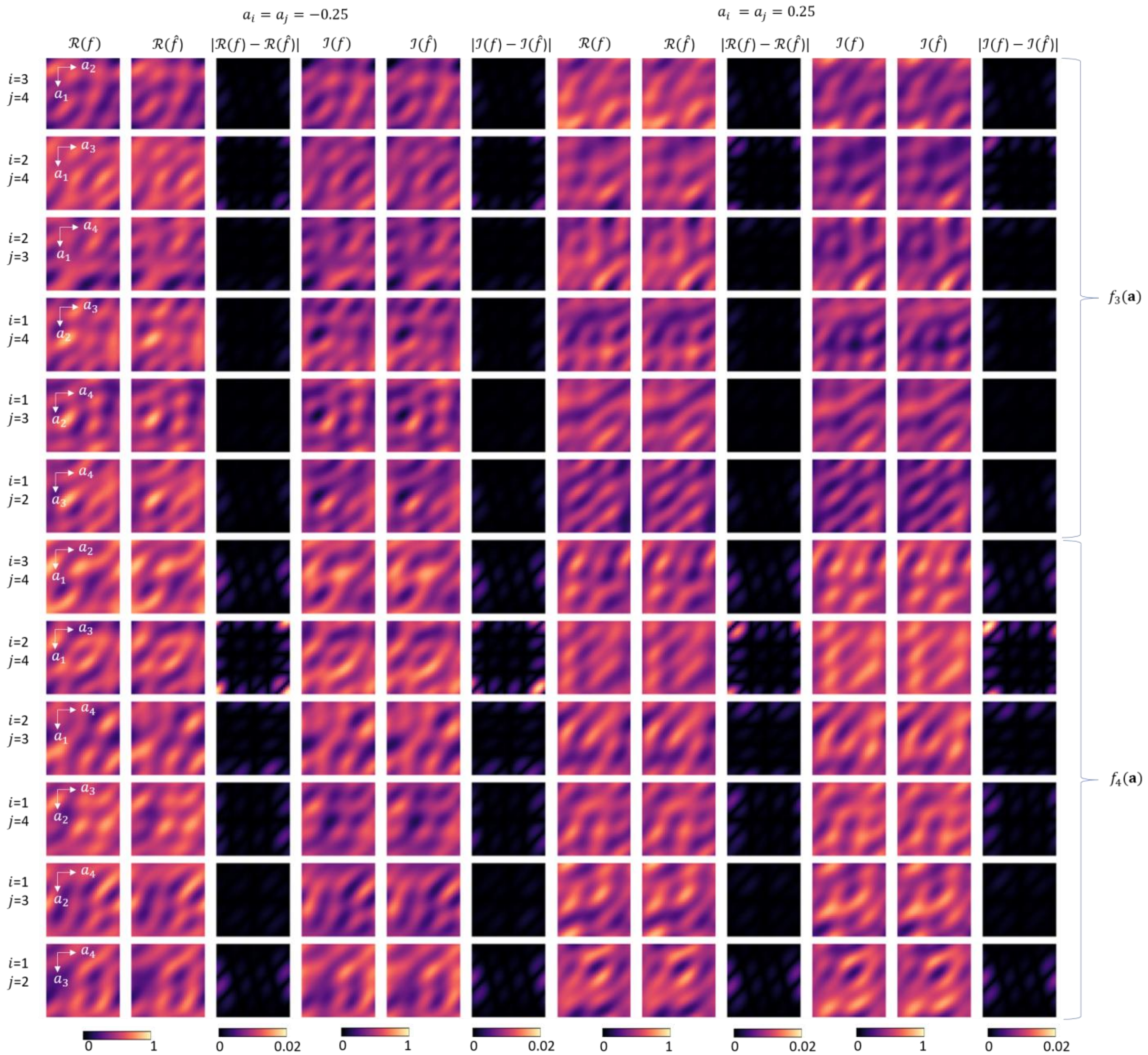

**Fig. 8 | 2-D slices of complex-valued 4-D nonlinear functions, comparing the ground truth nonlinear functions $f_3(a_1, a_2, a_3, a_4)$ and $f_4(a_1, a_2, a_3, a_4)$ with their all-optical approximations $\hat{f}_3(a_1, a_2, a_3, a_4)$ and $\hat{f}_4(a_1, a_2, a_3, a_4)$, respectively.** Each row shows one unique slice, where two coordinates are fixed to a common value, and the remaining two coordinates form the displayed 2-D plane. Within each group, the panels show the target nonlinear function, diffractive processor output, and absolute error for both the real ($\mathcal{R}$) and imaginary ($\mathcal{I}$) parts of each function. $N_p = F^D = 4^4 = 256$, and $K_L = 8$; see Fig. 6. The 4-D MAE values, computed over the full 4-D validation grid, are 0.001364 for $f_3$ and 0.003287 for $f_4$.

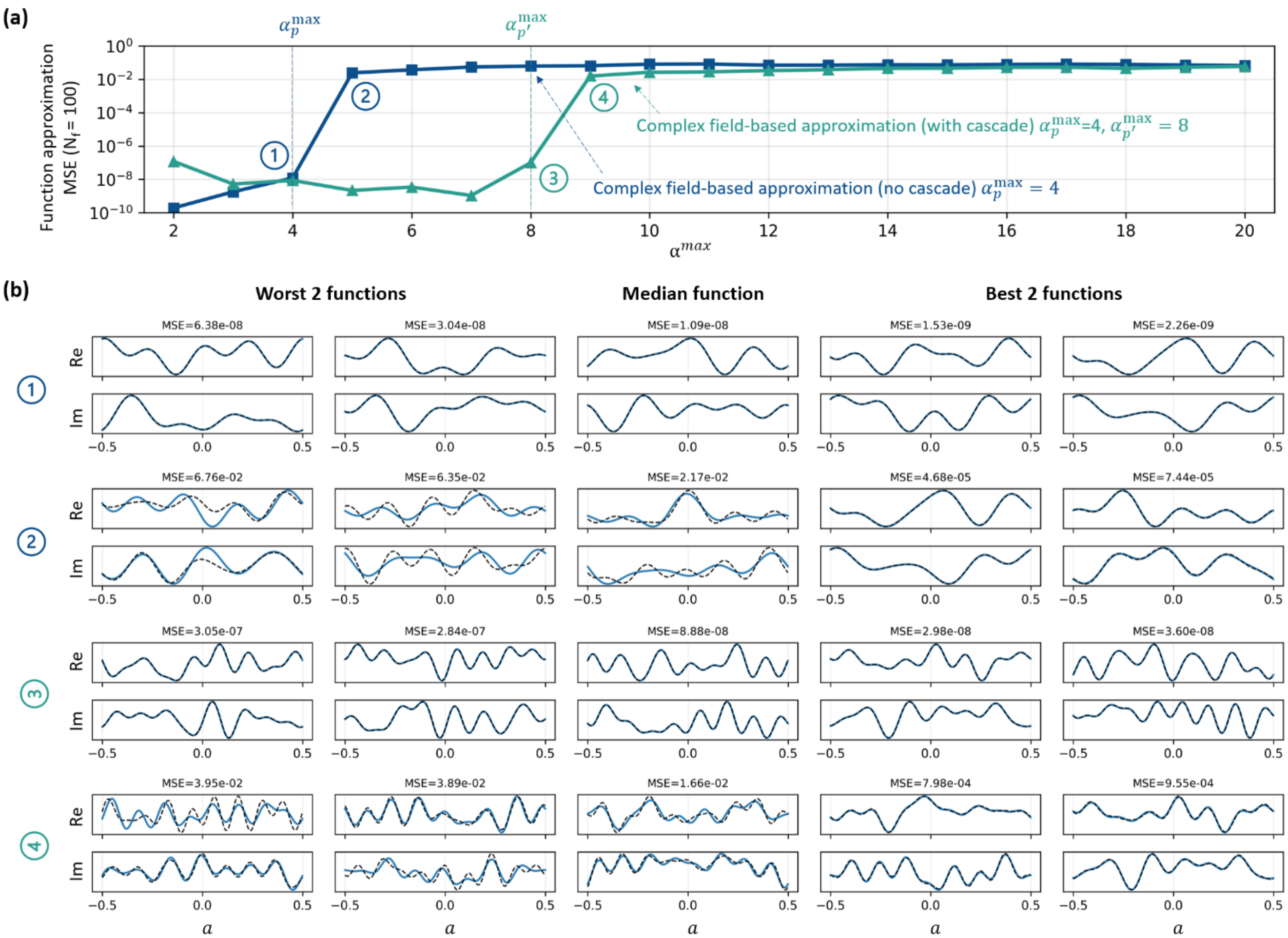

**Fig. 9 | All-optical cascadability of diffractive processors for optical field-based complex-valued nonlinear function approximation with additional intermediate input encoder channels ($p'$).** In this analysis, the 1st diffractive processor uses $N_p = 9$ input encoder channels to simultaneously approximate $N_f = 100$ complex-valued nonlinear functions at its output. It contains $N \approx 1.25 \times 2N_pN_f$ trainable phase-only features distributed over $K_L = 4$ diffractive surfaces. The 2nd diffractive processor jointly receives the $N_f = 100$ optical field outputs of the 1st processor as well as $N_{p'} = 8$ additional intermediate input encoder channels, and it is optimized to approximate $N_{f,2} = 100$ target nonlinear functions at its output plane. This 2nd cascaded diffractive processor contains $N \approx 1.25 \times 2(N_p + N_{p'})N_{f,2}$ trainable phase-only features distributed over $K_L = 4$ diffractive surfaces. **(a)** Mean normalized MSE of the optical field-based approximated complex-valued nonlinear functions vs. the target functions' maximum frequency $\alpha^{max}$. The circled markers (1–4) indicate representative functions highlighted in (b), covering the worst, best and median MSE values. **(b)** Representative examples (based on the worst, best and median MSE) of the optical field-based approximated nonlinear functions corresponding to the circled points in (a). Solid blue curves denote the diffractive approximations and dashed black curves denote the corresponding ground-truth target

functions; real and imaginary parts are separately shown for each complex-valued function for an input argument range of $a = [-0.5 : 0.5]$.

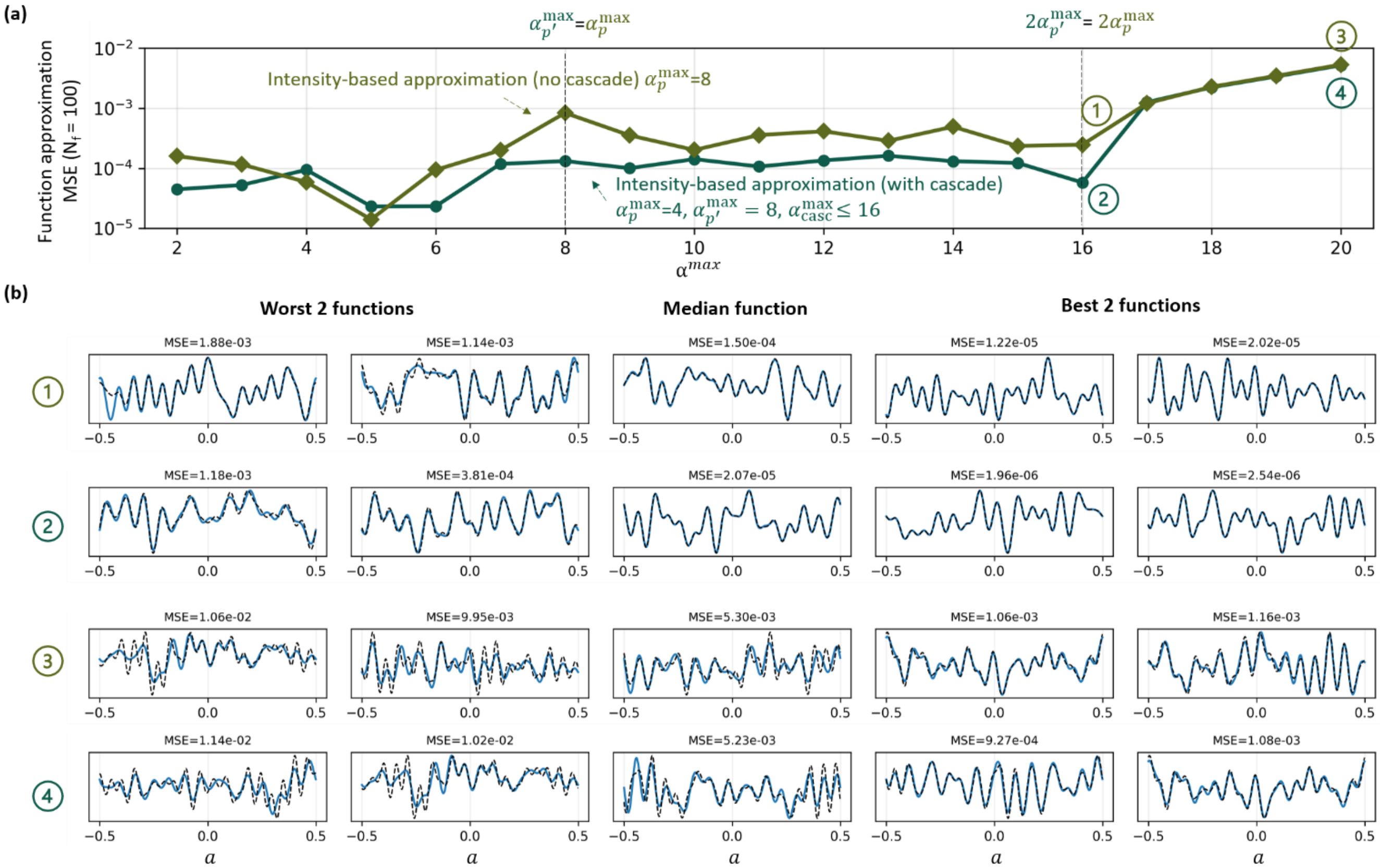

**Fig. 10 | All-optical cascadability of diffractive processors for optical intensity-based nonlinear function approximation with additional intermediate input encoder channels ($p'$).** The cascaded diffractive architecture has the same structural dimensions as in Fig. 9, including the same numbers of diffractive layers and trainable phase-only features in each diffractive processor. For comparison, the single diffractive processor (marked "no cascade" in (a)) uses $N_p = 17$ input channels and $N_{f,2} = 100$ output functions, with $N = 1.25 \times 2N_pN_{f,2}$ trainable phase-only features distributed over $K_L = 4$ diffractive surfaces. **(a)** Mean normalized MSE of the approximated intensity-based nonlinear functions vs. the target functions' maximum frequency $\alpha^{\max}$. This comparison is paired for each $\alpha^{\max}$ value, i.e., the same 100 distinct target nonlinear functions separately generated for each value of $\alpha^{\max}$ are used to compare the two diffractive architectures. The circled markers (1–4) indicate representative functions highlighted in (b), covering the worst, best and median MSE values. **(b)** Representative examples (based on the worst, best and median MSE) of the intensity-based approximated nonlinear functions corresponding to the circled points in (a). Solid blue curves denote the diffractive function approximations, and dashed black curves denote the corresponding ground-truth target functions.